\documentclass[journal]{IEEEtran}
\usepackage{amsmath, amssymb, amsthm, mathtools, stmaryrd, bm, gensymb}
\usepackage{algorithm, algpseudocode}
\usepackage{siunitx}
\usepackage{graphicx, subcaption}
\usepackage{optidef}
\usepackage{tikz, pgfplots}

\usetikzlibrary{fit,shapes}

\AtBeginEnvironment{thebibliography}{\linespread{0.96}\selectfont}

\makeatletter
\renewcommand{\ALG@beginalgorithmic}{\footnotesize}
\makeatother

\DeclarePairedDelimiterX{\openinterval}[2]{]}{]}{#1,#2}
\DeclarePairedDelimiterX{\openintervalboth}[2]{]}{[}{#1,#2}
\DeclarePairedDelimiterX{\closedinterval}[2]{[}{]}{#1,#2}

\mathtoolsset{showonlyrefs}

\def\tablename{Table}

\newcommand{\mb}[1]{\ensuremath{\mathbf{#1}}}				
\newcommand{\mc}[1]{\ensuremath{\mathcal{#1}}}				
\newcommand{\mbb}[1]{\ensuremath{\mathbb{#1}}}				
\newcommand{\tran}{\mathsf{T}}						
\newcommand{\hermit}{\mathsf{H}}					
\newcommand{\frob}[1]{\ensuremath{\left\|#1\right\|_\textrm{F}}}	
\newcommand{\esp}[1]{\ensuremath{\mathbb{E}\left[#1\right]}}		
\newcommand{\Fbb}{\mb{F}_{\text{BB}}}
\newcommand{\Frf}{\mb{F}_{\text{RF}}}

\newcommand{\sFrf}{\mc{F}_\text{RF}}
\newcommand{\Qb}{Q_b}
\newcommand{\vQb}{\mc{Q}_b}

\newcommand{\Heq}{\mb{H}_{\text{eq}}}

\newcommand{\wfsqrt}{\mb{\Lambda}^{1/2}}
\newcommand{\wf}{\mb{\Lambda}}
\newcommand{\Prf}{P_{\text{RF}}}
\newcommand{\Ups}{\mb{\Upsilon}_b}
\newcommand{\tU}{\mb{U}}
\newcommand{\tS}{\mb{\Sigma}}
\newcommand{\tV}{\mb{V}}

\newcommand{\Pmax}{P_\text{max}}
\newcommand{\Lrf}{L_\text{RF}}

\DeclareMathOperator{\diag}{diag}
\DeclareMathOperator{\Diag}{Diag}
\DeclareMathOperator{\Diagblk}{Diagblk}
\DeclareMathOperator{\tr}{tr}

\begin{document}
	
	\title{Energy Efficiency of mmWave Massive MIMO Precoding with Low-Resolution DACs}
	
	\author{Lucas~N.~Ribeiro,~\IEEEmembership{Student Member,~IEEE},
			Stefan~Schwarz,~\IEEEmembership{Member,~IEEE}, 
			Markus~Rupp,~\IEEEmembership{Fellow,~IEEE},
			Andr\'e~L.~F.~de~Almeida,~\IEEEmembership{Senior Member,~IEEE}
			
			\thanks{
				This work was supported in parts by the Erasmus Mundus SMART2 program, the Austrian Federal Ministry of	Science, Research and Economy and the National Foundation for Research, Technology and Development, and the Brazilian Coordination of Improvement of Higher Level Personnel (CAPES).
				
				The research has been  co-financed  by  the  Czech  Science  Foundation,  Project  No. 17-18675S ``Future  transceiver  techniques  for  the  society  in motion,''  and  by  the  Czech Ministry  of  Education  in  the  frame of  the  National  Sustainability  Program under grant LO1401.			
				
				L.~N.~Ribeiro, S.~Schwarz, and M.~Rupp are with the Institute of Telecommunications, TU Wien, 1040 Vienna, Austria. S. Schwarz is also member of the Christian Doppler Laboratory for Dependable Wireless Connectivity for the Society in Motion, TU Wien (e-mail: \{lribeiro, sschwarz, mrupp\}@nt.tuwien.ac.at).
				
				L.~N.~Ribeiro and A.~L.~F.~de~Almeida are with the Wireless Telecommunications Research Group (GTEL), Federal University of Cear\'a, Fortaleza, Brazil (email: \{nogueira, andre\}@gtel.ufc.br).
			}
		}
	
	\IEEEpubid{\begin{minipage}{\textwidth}\ \\[12pt] \centering
			 DOI: 10.1109/JSTSP.2018.2824762  \copyright 2018 IEEE. Personal use is permitted, but republication/redistribution requires IEEE permission.\\
			See http://www.ieee.org/publications\_standards/publications/rights/index.html for more information.
	\end{minipage}} 
	
	\maketitle
	
	\begin{abstract}
		With the congestion of the sub-$6\,\si{\giga\hertz}$ spectrum, the interest in massive multiple-input multiple-output (MIMO) systems operating on millimeter wave spectrum grows. In order to reduce the power consumption of such massive MIMO systems, hybrid analog/digital transceivers and application of low-resolution digital-to-analog/analog-to-digital converters have been recently proposed. In this work, we investigate the energy efficiency of quantized hybrid transmitters equipped with a fully/partially-connected phase-shifting network composed of active/passive phase-shifters and compare it to that of quantized digital precoders. We introduce a quantized single-user MIMO system model based on an additive quantization noise approximation considering realistic power consumption and loss models to evaluate the spectral and energy efficiencies of the transmit precoding methods. Simulation results show that partially-connected hybrid precoders can be more energy-efficient compared to digital precoders, while fully-connected hybrid precoders exhibit poor energy efficiency in general. Also, the topology of phase-shifting components offers an energy-spectral efficiency trade-off: active phase-shifters provide higher data rates, while passive phase-shifters maintain better energy efficiency.
	\end{abstract}

		
	\IEEEpeerreviewmaketitle
	
	\section{Introduction}
	
	\IEEEPARstart{T}{he} rapid increase of connected mobile terminals in the past few years has been pushing data rate requirements of 5G systems to new levels \cite{wang_cellular_2014}. As the sub-$6\,\si{\giga\hertz}$ spectrum is congested,  moving towards other ranges, such as millimeter wave (mmWave) regime, has been one of the main ideas for achieving these requirements \cite{andrews_what_2014}. The mmWave band is little regulated and is mostly available, allowing for mobile communication systems to operate on large bandwidths. However, the propagation characteristics on this elevated frequency range poses many engineering challenges.  For instance, according to Friis' Law, the isotropic path loss in free-space propagation is inversely proportional to the wavelength squared~\cite{constantine2005antenna}. It implies that path loss at the mmWave range is more severe than in lower frequency ranges in general. Nevertheless, for a given physical aperture, the antenna directional gain is also inversely proportional to the wavelength squared, and, thus, employing an array of highly directional antennas more than compensates for the free-space path loss~\cite{heath_overview_2016}. Such technology has been referred to in literature as mmWave massive multiple-input multiple-output (MIMO)~\cite{larsson_massive_2014, schwarz_society_2016, heath_millimeter_2016}.
	
	Hybrid analog/digital (A/D) transceiver architectures have been proposed to enable mmWave massive MIMO systems~\cite{heath_millimeter_2016}. They employ digital filtering (precoding/decoding) at baseband, and perform beamforming in the radio-frequency (RF) domain by analog components. The most popular implementation of this RF beamformer consists of an active phase-shifting network (PSN) connecting the outputs of the baseband filter to the antennas, which is known as fully-connected PSN. This implementation, however, is associated with a large power consumption as a considerable number of active phase-shift elements is required. As an alternative, one can employ sub-array beamforming, reducing the number of phase-shifters, and, consequently, power consumption. In this case, the PSN is said to be partially-connected. It has been claimed that hybrid precoding provides a data throughput close to that of fully-digital systems \cite{el_ayach_spatially_2014}. However, insertion losses of RF hardware are usually disregarded in the analysis of such hybrid systems. If these losses are not properly compensated for, then their spectral efficiency might be much smaller in practice than what is expected.
	
	Another energy-efficient approach to mmWave massive MIMO consists of using low-resolution digital-to-analog/analog-to-digital converters (DACs/ADCs) \cite{heath_overview_2016}. At the receive side, the high-resolution ADC chains are the most power hungry part, motivating the application of low-resolution devices to reduce their power consumption. At the transmit side, however, power expenditure is dominated by power amplifiers (PAs), which are usually required to operate within the high linearity regime to avoid distortion of the signal constellation. Employing low-resolution DACs relaxes the linearity requirement, allowing the amplifiers to operate closer to saturation, thus increasing their efficiency.
	
	\IEEEpubidadjcol
	
	\subsection{Related Work}

	Most of recent works on massive MIMO focus on analyzing either the performance of energy-efficient full-resolution hybrid or low-resolution fully-digital transceivers. In the following, we discuss their contributions.

	\subsubsection{Energy Efficiency of Hybrid Systems}
	
	The work of~\cite{mendez-rial_hybrid_2016} investigates the spectral and energy efficiencies of hybrid systems with switches and phase-shifters to perform analog beamforming. In order to evaluate energy efficiency, they define a power consumption model for both types of RF beamforming considering different interconnection of components. They also propose a channel estimation technique based on compressive sensing. According to simulation results, all hybrid architectures yield similar spectral efficiency for a given power consumption. In \cite{tsinos_energy-efficiency_2017}, the energetic performance of single/multi-carrier full-resolution hybrid transceivers is investigated. A transceiver optimization problem based on energy efficiency maximization is presented and solved by the alternating direction of multipliers method. The power consumption model proposed therein considers the computational power expenditure and RF hardware losses, which are usually ignored in most energy efficiency models. This model assumes the application of PAs and low-noise amplifiers that compensate for the analog beamforming losses. Based on this assumption, the paper claims that hybrid precoders with fully-connected PSN can be energy-efficient when the transmit power is high, even more than those with partially-connected PSN. This conclusion, however, contradicts the results of \cite{garcia-rodriguez_hybrid_2016}, where the spectral efficiency of a hybrid precoder is examined under a realistic RF model. Therein, the fully-connected PSN is modeled as a bank of RF components described by their S-parameters. The obtained results show that signal-to-noise ratio (SNR) losses are significant, going up to $25\,\si{\decibel}$ for the given scenario. Unfortunately, with the current mmWave amplifier technology, one cannot assume that these losses are simply compensated for as in \cite{tsinos_energy-efficiency_2017}. Therefore, in order to make a more realistic comparison between hybrid and digital transmitters, the effect of RF losses on the spectral efficiency has to be considered.
	
	Analog and digital receivers employing low-resolution ADCs are studied in \cite{orhan_low_2015} for single-user MIMO (SU-MIMO). The authors resort to a stochastic linear quantization model referred to as additive quantization noise (AQN) model in order to simplify the analysis of systems with low-resolution ADCs. For digital combining, singular value decomposition (SVD) processing with water-filling power allocation is applied, whereas much simpler matched filtering is employed in analog beamforming. Lower bounds on data rate are derived based on the linear quantization approximation, and the AQN model is shown to be accurate in the low SNR regime. Simulation results indicate that digital combining exhibits better performance than the analog scheme. The energy and spectral efficiency trade-off for digital, analog, and hybrid receivers is extensively studied in \cite{abbas_millimeter_2016}. The ADC is approximated by the AQN model and achievable rate expressions are obtained. A power consumption model similar to that of \cite{mendez-rial_hybrid_2016} is considered, allowing to calculate energy efficiency. At the transmit side, the authors assume fully-digital precoding, whereas, at the receive side, fully-connected RF beamforming is employed in addition to digital baseband combining. The analog beamformers are computed by the alternating minimization method of \cite{tropp_designing_2005}, and the baseband filter is obtained through SVD processing with water-filling power allocation. Results indicate that analog combining is the most energy-efficient solution only at low SNR or low-rank channels, while the efficiency of hybrid and digital combining strongly depends on the assumed hardware power consumption characteristics for any other than low-rank channels. It is important to stress that \cite{abbas_millimeter_2016} does not consider insertion losses at the RF domain and only investigates systems with fully-connected PSN. In \cite{roth_achievable_2016}, by contrast, hybrid receivers with partially-connected PSN are compared to digital combiners under a low-resolution ADC assumption. This contribution shows that digital receivers using low-resolution ADCs are robust to small automatic gain control (AGC) imperfections. In the considered power consumption model, a simple implementation for a $1$-bit ADC is presented, leading to negligible power consumption. The considered analog combining strategy relies on beam scanning, and baseband combining consists of SVD processing with water-filling power allocation, as in the previously mentioned works. Results suggest that digital combining is more efficient than hybrid combining especially in the low SNR regime. However, the contributions mentioned above impose low-resolution only at the receiver side, assuming fully-digital or hybrid transmitters with high-resolution DACs.
	
	\subsubsection{Transmitters with Low-Resolution DACs}
	
	Recently proposed massive MIMO signal processing methods such as~\cite{marzetta_fundamentals_2016,ribeiro_low-complexity_2017} assume very large antenna arrays with dedicated RF chains either at the transmit or at the receive side. However, such implementation is not energy-efficient and thus needs to be modified in order to reduce its power requirements. An alternative to hybrid systems consists of employing low-resolution DACs/ADCs at transmitter/receiver to relax the power demands on the fully-digital transceiver. 
	
	A SU-MIMO model with low-resolution quantization at the transmit side is introduced in \cite{mezghani_transmit_2009}. A linear approximation for DAC quantization based on the Bussgang Theorem~\cite{bussgang_crosscorrelation_1952} is presented, allowing the derivation of a minimum mean square error (MMSE) precoder optimized for tackling the quantization effects. Bit error ratio results reveal that this optimized MMSE filter outperforms the plain MMSE solution. In \cite{jacobsson_quantized_2016}, a narrow-band multi-user MIMO (MU-MIMO) system employing low-resolution DACs at the base station is considered. The authors investigate the performance of linear precoders with coarse quantization and propose a variety of non-linear precoders based on relaxations of the MMSE-optimal downlink precoding problem. Achievable rate expressions are obtained and simulation results suggest that performance achieved with infinite-resolution DAC can be attained by using $3$ or $4$ bits of resolution for the given scenario. Furthermore, it was shown that the presented non-linear precoding algorithms significantly outperform the linear filtering solutions for $1$-bit quantization.
	
	\subsection{Contribution}
	
	To the best of our knowledge, extensive performance evaluation of quantized hybrid transmitters with fully- and partially-connected PSN under realistic RF modeling has not been considered yet. We thus aim at filling this gap by this paper. Our main contributions in this direction can be summarized as follows:
	\begin{itemize}
		\item A SU-MIMO system model for quantized hybrid precoding based on the AQN model is presented. The proposed system model differs from those previously introduced in other works in the definition of the total additive noise vector, which accounts for the additive white Gaussian noise (AWGN) and an RF-filtered quantization noise.
		\item We define the quantized hybrid precoding problem, which can be regarded as a generalization of the quantized digital precoding and classical hybrid precoding problems. Achievable rate expressions for both schemes are derived. For hybrid precoding, we consider both fully- and partially-connected PSNs. The effects of DAC quantization on spectral efficiency are assessed through asymptotic analysis.
		\item A power consumption model considering device power characteristics reported in recent mmWave literature is presented. Active and passive phase-shifting topologies are considered in our model. The former has important power consumption and amplifies the shifted signals, whereas the latter has insignificant power consumption and substantial  insertion losses. We properly account for power insertion losses in our signal model using a realistic RF modeling. The effects of signal processing computations are also considered in the power budget by the model presented in \cite{bjornson_optimal_2015}. Simulations are conducted to evaluate the importance of this computational term and the obtained results suggest that it may be crucial for large-array systems.
		\item We introduce an analog beamforming method for partially-connected structures based on maximum eigenmode transmission (MET), and implement this solution with the low-complexity power method of \cite[Section 7.3.1]{golub_matrix_2012}. The computational complexity of the precoding methods are determined in order to calculate their power consumption.
		\item The energy-spectral efficiency trade-off is investigated to determine the most efficient precoding strategy, and to study the influence of phase-shifter implementation on performance. Our results show that hybrid precoding with partially-connected PSN and digital precoding are the most energy- and spectral-efficient solutions, respectively. Fully-connected PSN, however, exhibits poor energetic performance due to substantial power consumption and large insertion losses, which cause severe spectral efficiency degradation. Moreover, we observed that active phase-shifters favor spectral efficiency, while their passive implementations focus on energy efficiency.
	\end{itemize}

	This work is organized as follows: Section~\ref{sec:sysmodel} describes the system model considered in this work. Signal and channel models for a narrow-band mmWave SU-MIMO system are introduced therein. The quantization operation employed in DACs is defined and approximated by the AQN model, and the power consumption and loss models are presented. The analog and digital precoding strategies adopted in this work are defined in Section~\ref{sec:precstr}. Also, computational complexity analysis is conducted in Section~\ref{sec:companalysis} and achievable rate bounds are derived in Section~\ref{sec:achrate}. The proposed models and precoding methods are assessed through simulations in Section~\ref{sec:sim}, and the work is concluded in Section~\ref{sec:conc}.

	\subsection{Notation}
	
	In this paper, $x$ denotes a scalar, $\mb{x}$ a vector, and $\mb{X}$ a matrix. The $(i,j)$-th entry of $\mb{X}$ is given by $[\mb{X}]_{i,j}$. The transposed and conjugated transposed (Hermitian) of $\mb{X}$ are denoted by $\mb{X}^\tran$ and $\mb{X}^\hermit$, respectively. The $(M\times M)$-dimensional identity matrix is represented by $\mb{I}_M$. The $(M \times N)$-dimensional null matrix is given by $\mb{0}_{M \times N}$. $\tr(\cdot)$ is the matrix trace, $\diag(\cdot)$ forms a diagonal matrix out of the main diagonal of the matrix argument, $\Diag(\cdot)$ transforms the input vector into a diagonal matrix, $\Diagblk(\cdot)$ forms a block-diagonal matrix from the matrix inputs, and $\det(\cdot)$ is the determinant. The operator $\angle(\cdot)$ extracts the angle from the argument's complex elements. The absolute value, the Frobenius and $\ell_2$ norms, and the expected value operator are respectively represented by $|\cdot|$, $\frob{\cdot}$, $\| \cdot \|_2$, and $\esp{\cdot}$. The operators $\Qb(\cdot)$ and $\vQb(\cdot)$ denote scalar and vector quantization, respectively. The scalar ceiling function is given by $\lceil \cdot \rceil$. The set of circularly-symmetric jointly-Gaussian complex random vectors with mean vector $\bm{\mu}$ and covariance matrix $\mb{C}$ is denoted by $\mc{CN}(\bm{\mu}, \mb{C})$.
	
	\section{System Model} \label{sec:sysmodel}
	
	\subsection{Signal Model}
	
	Consider a single-user mmWave MIMO system in which the transmitter sends $N_s$ streams of data symbols through an array of $N_t$ antennas to the receiver equipped with $N_r$ antennas. The discrete-time data streams, represented by the vector $\mb{s}=[s_1, \ldots, s_{N_s}]^\tran$, are assumed to be independent and Gaussian distributed with zero mean and unit variance, hence their covariance matrix is $\mb{R}_{ss} = \mbb{E}[\mb{s}\mb{s}^\hermit] = \mb{I}_{N_s}$. Before transmission, the data streams are precoded as  $\mb{x} = \mc{P}(\mb{s}) \in \mbb{C}^{N_t}$ satisfying the average power constraint $P_x = \esp{\| \mb{x}\|_2^2} \leq \Pmax$, where $\mc{P}(\cdot):\mbb{C}^{N_s} \rightarrow \mbb{C}^{N_t}$ represents the precoding operator.  The propagation channel is assumed to be narrow-band block-fading, so there is no inter-symbol interference, and the discrete-time representation of the received signal can be expressed as
	\begin{equation} \label{eq:recv_sig}
		\mb{y} = \mb{Hx} + \mb{n} \in \mbb{C}^{N_r},
	\end{equation}
	where $\mb{H} \in \mbb{C}^{N_r \times N_t}$ denotes the MIMO channel matrix, and $\mb{n} \sim \mc{CN}(\mb{0}_{N_r \times 1}, \sigma_n^2 \mb{I}_{N_r})$ is the AWGN vector. The SNR normalized with respect to the maximum average transmit power is defined as $\gamma = \Pmax/\sigma_n^2$. 
	
\begin{figure}
		\centering
		\begin{subfigure}[b]{0.4\textwidth}
			\includegraphics[width=\textwidth]{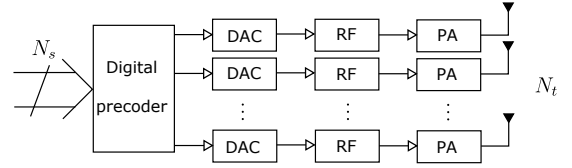}
			\caption{Low-resolution digital precoder architecture.}
			\label{fig:lowresdig}
		\end{subfigure}
		\begin{subfigure}[b]{0.4\textwidth}
			\includegraphics[width=\textwidth]{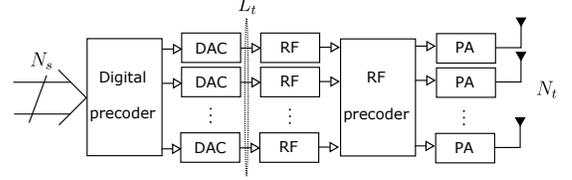}
			\caption{Low-resolution hybrid precoder with $L_t$ RF chains, and fully-connected PSN.}
			\label{fig:hpfull}
		\end{subfigure}
		\begin{subfigure}[b]{0.4\textwidth}
		\includegraphics[width=\textwidth]{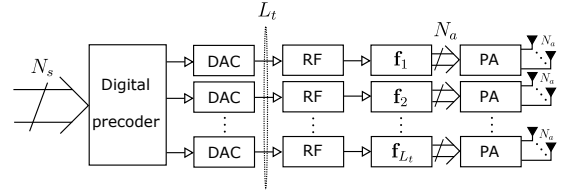}
		\caption{Low-resolution hybrid precoder with $L_t$ RF chains, partially-connected PSN, and $L_t$ sub-arrays of $N_a$ antennas.}
		\label{fig:hppart}
		\end{subfigure}
		\caption{Transmitter architectures.}\label{fig:animals}
	\end{figure}

	We consider digital and hybrid A/D precoding schemes. The former consists of precoding the data streams using only a digital baseband filter $\Fbb \in \mbb{C}^{N_t \times N_s}$, which feeds signals to $N_t$ DAC/RF chain pairs connected PAs and then to the antenna array, as illustrated in Figure~\ref{fig:lowresdig}. The latter scheme aims at decreasing the precoding power consumption by employing only $N_s \leq L_t < N_t$ DAC/RF chain pairs. To achieve this, the $N_s$ data streams are first precoded by a digital baseband precoder $\Fbb \in \mbb{C}^{L_t \times N_s}$, which feeds $L_t$ DAC/RF chain pairs. Next, the up-converted signals are filtered by an analog precoder, amplified and transmitted through the antenna array. We consider that the analog beamforming is implemented by low-resolution phase-shifters as in~\cite{sohrabi_hybrid_2015}. Fully- and partially-connected PSN topologies are considered for analog beamforming. In fully-connected PSNs, the output of each RF chain is split by a $N_t$-way power divider and connected to $N_t$ phase-shifters. Afterwards, the shifted signals are merged by $L_t$-way power combiners at each PA/antenna pair. In this case, the analog precoder is represented by a full matrix $\Frf \in \mbb{C}^{N_t \times L_t}$ whose entries are constrained to have constant modulus and discrete phase resolution, i.e., $\left[ \Frf \right]_{m,n} \in \sFrf,\, m \in \{1,\ldots, N_t\},\,n \in \{1,\ldots,L_t\}$. The set $\mc{F}_{\text{RF}} = \{1, \phi, \phi^2, \ldots, \phi^{n_\text{PS}-1}\}$ denotes the phase-shift set, $\phi = \exp(j2\pi/n_\text{PS})$ the angular resolution, $n_\text{PS} = 2^{b_\text{PS}}$ the number of phase-shifts supported by the hardware, and $b_\text{PS}$ the bit resolution. This architecture employs in total $L_t N_t$ phase-shift elements and is illustrated in Figure~\ref{fig:hpfull}. In partially-connected PSNs, by contrast, each RF chain is linked to only a sub-array of $N_a = \lceil N_t/L_t \rceil$ phase-shifters directly connected to PA/antenna pairs, as illustrated in Figure~\ref{fig:hppart}. Therefore, the PSN for this architecture employs only $N_a$-way power dividers and no power combiners at all. The analog beamforming matrix in this case presents the following structure: $\Frf = \Diagblk(\mb{f}_1, \ldots, \mb{f}_{L_t}) \in \mbb{C}^{N_t \times L_t}$, where $\mb{f}_\ell \in \mbb{C}^{N_a}$ denotes the analog beamforming filter of the $\ell$-th sub-array. The phase-shifters constrain the sub-arrays analog beamforming filters as $\left[ \mb{f}_\ell \right]_{m} \in \mc{F}_\text{RF},\,\ell = 1,\ldots,L_t, \,m \in \{1,\ldots, N_a\}$.
	
	To make our model consistent with power constraints present in real-world implementations, we incorporate non-linearities due to the DAC stages, and RF losses caused by analog beamforming. The non-linearity introduced by DACs is modeled as a quantization stage. This is because the continuous-valued DAC output signal has its amplitude well-represented by a finite set of values generated by the DAC hold circuits. The RF losses are accounted by considering a power loss factor $\frac{1}{\sqrt{\Lrf}}$ multiplying the analog beamforming matrix $\Frf$, as in \cite{garcia-rodriguez_hybrid_2016}. The insertion loss of the analog precoder is represented by $\Lrf$ and depends on the power characteristics of the dividers, combiners and phase-shifters. The power loss model is further detailed in Section~\ref{sec:pow}. To include DAC and RF losses in our model, we define the precoding operation as
	\begin{equation} \label{eq:precoding_model}
		\mb{x} = \tfrac{1}{\sqrt{\Lrf}} \Frf \vQb( \Fbb \mb{s} )  =\tfrac{1}{\sqrt{\Lrf}} \tilde{ \mb{x} },
	\end{equation}
	where $\vQb(\cdot)$ stands for the vector quantization operator with $b$ bits of resolution per dimension, and $\tilde{ \mb{x} } = \Frf \vQb( \Fbb \mb{s} )$ the lossless transmitted signal. The vector quantization operator will be further detailed and linearly approximated in Section~\ref{sec:quant_model}. Notice that~\eqref{eq:precoding_model} refers to digital precoding by setting $\Frf = \mb{I}_{N_t}$, $\Lrf=1$, and $\Fbb \in \mbb{C}^{N_t \times N_s}$. Therefore, for notation convenience, we consider that the quantization input and output vectors are $M$-dimensional, where $M=N_t$ and $M=L_t$ for digital and hybrid precoding, respectively. Thus, $\Frf$ and $\Fbb$ are $(N_t \times M)$ and $(M \times N_s)$-dimensional matrices, respectively.
	
	\subsection{Quantized Signal Model}	 \label{sec:quant_model}

	Let us define the quantization operation used in this work to model DAC and introduce its linear approximation. We define $\Qb(\cdot)$ as the uniform scalar quantizer that operates independently on both real and imaginary components of the input signal. A quantizer with $b$ resolution bits has $N_b=2^b$ codes for representing the quantized values. The quantizer output is $p = \Qb(u) = c_i^R + j c_\ell^I$ when $\mathfrak{Re}[u] \in\, \openinterval{a_{i-1}}{a_i}$ and $\mathfrak{Im}[u] \in \openinterval{b_{\ell-1}}{b_\ell}$, for $i,\ell \in \{1,\ldots,N_b-1\}$. The value $c_i^R$ ($c_\ell^I$) represents the code associated with the real (imaginary) part of the $i$-th ($\ell$-th) quantization interval, and $a_i$ ($b_\ell$) the quantization segment for the real (imaginary)  component. We set $a_0 = b_0 = -\infty$ and $a_{N_b-1} = b_{N_b-1} = \infty$ to support inputs with arbitrary power. The quantization codes are chosen to minimize the MSE of Gaussian distributed signals, and they are listed in \cite[Table II]{max_quantizing_1960}. Such coding suits our modeling since the baseband precoded signals are still Gaussian. For vector inputs, we define $\vQb(\cdot):\mbb{C}^{M} \rightarrow \mbb{C}^M$ such that $\mb{p} = \vQb(\mb{u}) \Rightarrow [\mb{p}]_m = \Qb([\mb{u}]_m),\quad m \in \{1,\ldots,M\}$. The vector quantizer simply consists of applying scalar quantization to each input vector entry. Unlike the Lloyd-Max quantizer \cite{makhoul_vector_1985}, the considered vector quantization approach is not optimal in the MSE sense.
	
	Conducting performance analysis of a MIMO system in terms of the exact non-linear quantization model presented above can be challenging. Among many reasons, we point to the difficulty of calculating the statistics of the transmitted signals. As an alternative, we resort to a linear approximation of the scalar quantizer output applying the AQN model \cite{orhan_low_2015,abbas_millimeter_2016,mezghani_capacity_2012} $p = \Qb(u) \approx \sqrt{1-\rho_b} u + e$. The input $u$ is assumed to be Gaussian distributed, $e$ is the quantization noise uncorrelated with $u$, and $\rho_b$ is a quantization distortion factor, which is defined as the ratio of the quantization noise variance $\sigma^2_e = \esp{|e|^2}$ to the input signal variance $\sigma_u^2 = \esp{|u|^2}$, i.e., $\rho_b =\sigma_e^2/\sigma_u^2$ \cite{mezghani_capacity_2012}. Values for this factor are listed in \cite[Table II, page 12]{max_quantizing_1960} for resolution up to $5$ bits. For larger resolution, it can be approximated as $\rho_b \approx \frac{\pi\sqrt{3}}{2}2^{-2b}$~\cite{mezghani_capacity_2012}. Note that the quantization distortion factor also represents the inverse of the signal-to-quantization-noise ratio (SQNR). Thus, when the SQNR goes to infinity, the bit resolution $b$ increases, and $\rho_b$ converges to zero. In this quantization model, the additive noise power is proportional to the input variance  and decreases as the quantization resolution grows. Such a behavior is coherent with practice and motivates the adoption of this model.  Also note that the approximation definition ensures that $\sigma_p^2 \approx \sigma_u^2$. Let us extend the AQN model to MIMO systems:
	\begin{equation} \label{eq:aqn}
	\mb{p} = \vQb(\mb{u}) \approx \Ups \mb{u} + \mb{e} \in \mbb{C}^M.
	\end{equation}
	Vectors $\mb{u} = [u_1,\ldots, u_M]^\tran$ and $\mb{p} = [p_1,\ldots, p_M]^\tran$ denote respectively the input and output of the quantizer, and $\mb{e} = [e_1, \ldots, e_M]^\tran$ the noise quantization vector which satisfies $\mbb{E}[\mb{u}\mb{e}^\hermit] = \mbb{E}[\mb{e}\mb{u}^\hermit] = \mb{0}_{M \times M}$. The diagonal matrix $\Ups = \Diag(\sqrt{1-\rho_{b,1}}, \ldots, \sqrt{1-\rho_{b,M}})$ is formed by the quantization distortion factors $\rho_{b,m}$ associated with the $m$-th quantizer input-output. From the definition of $\rho_b$, it follows that the covariance matrix of the quantization error vector is given by:
	\begin{equation} \label{eq:quant_error_cov_mtx}
	\mb{R}_{ee} = \Diag(\rho_{b, 1},\ldots,\rho_{b,M}) \diag(\mb{R}_{uu})\in \mbb{C}^{M \times M},
	\end{equation}
	where $\mb{R}_{uu}=\esp{\mb{u}\mb{u}^\hermit}$ denotes the quantizer input covariance matrix. Knowledge of these covariance matrices will be useful in Section~\ref{sec:precstr}, where we will derive the achievable rate for each precoder.
	
	The AQN model approximates a non-linear deterministic operation as a linear stochastic process by assuming zero correlation between the quantizer input and the quantization noise. Similar  results are obtained by using the Bussgang Theorem \cite{bussgang_crosscorrelation_1952}, which states that the cross-correlation function between two Gaussian signals taken after one of them has been non-linearly distorted is proportional to the cross-correlation function before distortion. If a linear model, e.g. \eqref{eq:aqn}, is used to approximate the non-linear distortion, this Theorem implies that the approximation error is uncorrelated with the input, i.e., $\mbb{E}[\mb{u}\mb{e}^\hermit] = \mbb{E}[\mb{e}\mb{u}^\hermit] = \mb{0}_{M \times M}$. Therefore, the Bussgang Theorem can be seen as the theoretical foundation underlying the AQN model, since this non-correlation assumption is crucial. It was shown in \cite{roth_achievable_2016} that AGC error could invalidate this assumption, but errors between $-20\%$ and $20\%$ are still acceptable. Model accuracy has been investigated in \cite{orhan_low_2015} and it was found to be accurate especially in the low SNR regime, provided that the input signals are Gaussian distributed. 
	
	Now let us apply the AQN model to the precoded signal model \eqref{eq:precoding_model}. Define the DAC input signal $\mb{u} = \Fbb \mb{s} \in \mbb{C}^M$. The lossless transmitted signal defined in \eqref{eq:precoding_model} can now be approximated as
	\begin{align} 
	\tilde{\mb{x}}  \approx \Frf \Ups\mb{u} +  \Frf \mb{e} \in \mbb{C}^{N_t}. \label{eq:precod_lowres}
	\end{align}
	Since the data streams $\{s_1, \ldots, s_{N_s}\}$ are Gaussian distributed, the baseband precoded signals $\{u_1, \ldots, u_{M}\}$ will also be Gaussian distributed with covariance matrix $\mb{R}_{uu} = \Fbb \Fbb^\hermit$. As each quantizer output is Gaussian distributed and has $b$ bits of resolution, it is reasonable to claim that all DACs undergo the same distortion  $\rho_b$, i.e., $\rho_{b,1}= \rho_{b,2}= \ldots = \rho_{b,M} = \rho_b$, and, consequently, it follows that $\Ups = \sqrt{1-\rho_b} \mb{I}_M$. Now the error covariance matrix \eqref{eq:quant_error_cov_mtx} simplifies to
	\begin{equation} 
	\mb{R}_{ee} = \rho_b \diag(\mb{R}_{uu}). \label{eq:cov_quant_error_lowres}
	\end{equation}
	Using \eqref{eq:precoding_model} and \eqref{eq:precod_lowres}, the received signal model \eqref{eq:recv_sig} can be approximated as:
	\begin{align}  
	\mb{y} =  \mb{H} \mb{x} + \mb{n} \approx \tfrac{1}{\sqrt{\Lrf}}\mb{H}' \mb{u} + \mb{n}',\label{eq:eqmimo}
	\end{align}
	where $\mb{H}' = \Heq \Ups$ denotes the $(N_r \times M)$-dimensional total channel matrix, $\Heq = \mb{H} \Frf \in \mbb{C}^{N_r \times M}$ the equivalent channel formed by the cascade of the analog precoding and the transmission channel, and $\mb{n}' = \frac{1}{\sqrt{\Lrf}} \Heq \mb{e} + \mb{n} \in \mbb{C}^{N_r}$ the total additive noise component. The total channel matrix $\mb{H}'$ is formed by the cascade of $\Heq$ and the quantization scaling $\Ups$. The former factor models the effect of the physical channel $\mb{H}$ and the lossless analog precoder $\Frf$ (for hybrid structures). Of course, one must have already designed $\Frf$ to form $\Heq$, and, indeed, computing $\Frf$ is the first step of the hybrid precoding strategies we present in Section~\ref{sec:precstr}. The factor $\Ups$, on the other hand, reflects the reduction of the useful transmit power due to quantization. For high-resolution quantization, $\rho_b \rightarrow 0$ and $\mb{H}' \rightarrow \Heq$, thus, in this case, the quantization noise has no impact on the total channel power. By contrast, for low- and mid-resolution quantization, $\Ups$ becomes important and the total channel is then expressed as $\mb{H}' = \sqrt{1-\rho_b} \Heq$. Since $\sqrt{1-\rho_b}$ is always smaller than one, low-resolution quantization decreases the total channel power. Another interesting feature of model \eqref{eq:eqmimo} is that $\mb{n}'$ comprises the AWGN plus an attenuated channel-filtered quantization noise term. Therefore, $\mb{n}'$ is not guaranteed to be Gaussian, as nothing is assumed about the distribution of $\mb{e}$. The covariance matrix of the total additive noise is
	\begin{equation} \label{eq:cov_totnoise}
	\mb{R}_{n'n'} = \mbb{E}[\mb{n}'\mb{n}^{\prime\hermit}] = \tfrac{1}{\Lrf}\Heq \mb{R}_{ee} \Heq^\hermit + \mb{R}_{nn} \in \mbb{C}^{N_r \times N_r},
	\end{equation}
	where $\mb{R}_{nn} = \sigma_n^2 \mb{I}_{N_r}$, and the quantization noise covariance matrix $\mb{R}_{ee}$ is given by Equation~\eqref{eq:cov_quant_error_lowres}. One can easily see that $\mb{R}_{n'n'}$ is in general not a diagonal matrix, therefore the total additive noise vector entries are correlated. Note that we assume $\esp{\mb{n}\mb{e}^\hermit} = \mb{0}_{N_r \times M}$, which is reasonable in practice, since the quantization process at the transmitter has no influence whatsoever on the receiver sensor noise.	
	
	Unfortunately, Equation~\eqref{eq:eqmimo} is still problematic when it comes to obtaining achievable rate expressions. It is difficult to separate the total MIMO channel $\mb{H}'$ into orthogonal subchannels due to the structure of the total noise vector $\mb{n}'$. There are no guarantees that $\mb{n}'$ is Gaussian distributed, complicating the achievable rate maximization. Moreover, its covariance matrix \eqref{eq:cov_totnoise} depends on $\mb{R}_{ee}$, which varies according to the diagonal elements of $\mb{R}_{uu}$, bringing a causality problem for the precoder design. In view of these difficulties, we make some assumptions to simplify our analysis and to obtain a lower bound for the achievable rate. In \cite{diggavi_worst_2001}, it was shown that the Gaussian noise distribution minimizes the mutual information for a given noise covariance matrix. This result motivates us to approximate the additive quantization noise $\mb{n}'$ as a jointly Gaussian distributed noise vector $\mb{n}_G = \frac{1}{\sqrt{\Lrf}}\Heq \mb{e} + \mb{n}$ with covariance matrix  $\mb{R}_{n_G n_G} = \mb{R}_{n' n'}$ in order to obtain an achievable rate lower bound as in \cite{mezghani_capacity_2012}. The received signal model~\eqref{eq:eqmimo} is now be approximated as 
	\begin{equation} \label{eq:approx_signal_ftw}
		\mb{y} \approx \frac{1}{\sqrt{\Lrf}}\mb{H}'\mb{u} + \mb{n}_G = \frac{1}{\sqrt{\Lrf}}\Heq \Ups \Fbb \mb{s} + \mb{n}_G.
	\end{equation}
	To simplify the analysis of this signal model, the colored noise vector $\mb{n}_G$ is decorrelated. Since $\mb{R}_{n_G n_G}$ is a Hermitian matrix, it admits an eigenvalue decomposition $\mb{R}_{n_G n_G}= \mb{J}\mb{L}\mb{J}^\hermit$, where $\mb{J}^\hermit \mb{J} = \mb{J}\mb{J}^\hermit = \mb{I}_{N_r}$. The whitening filter is thus given by $\mb{R}_{n_G n_G}^{-1/2} = \mb{L}^{-1/2}\mb{J}^\hermit$ \cite{yu_iterative_2004}, and the received signal vector $\mb{y}$ is pre-multiplied by this filter, yielding:
	\begin{equation} \label{eq:eqmimo_white}
	\mb{R}_{n_G n_G}^{-1/2} \mb{y} = \tfrac{1}{\sqrt{\Lrf}}\mb{R}_{n_G n_G}^{-1/2} \mb{H}' \mb{u} + \mb{R}_{n_G n_G}^{-1/2} \mb{n}_G.
	\end{equation} 
	Notice that after noise whitening, the covariance matrix of the noise vector in $\eqref{eq:eqmimo_white}$ is an $(N_r\times N_r)$-dimensional identity matrix. We further apply model \eqref{eq:eqmimo_white} in Section~\ref{sec:precstr} to define the quantized hybrid precoding problem.

	\subsection{Channel Model} \label{sec:channel}

	Experiments conducted in \cite{gao_mmwave_2015,zochmann_directional_2016} indicate that mmWave massive MIMO channels present a high degree of spatial and angular sparsity due to the large path loss, and thus there are only a few dominant multipaths. Such channels can be modeled by a narrow-band clustered channel model contributing with $L$ propagation paths \cite{heath_overview_2016,el_ayach_spatially_2014} resulting in a channel matrix
	\begin{equation} \label{eq:channel_vector}
	\mb{H} = \sqrt{\tfrac{N_t N_r}{L}}\sum_{\ell=1}^{L} \alpha_{\ell} \mb{a}_r(\theta_{\ell}^r, \phi_\ell^r) \mb{a}_t(\theta_{\ell}^t,\phi_\ell^t)^\hermit \in \mbb{C}^{N_r \times N_t},
	\end{equation}
	where $\alpha_{\ell} \in \mbb{C}$ denotes the small-scale fading, $\mb{a}_r(\theta_{\ell}^r, \phi_\ell^r) \in \mbb{C}^{N_r}$ and $\mb{a}_t(\theta_{\ell}^t,\phi_\ell^t) \in \mbb{C}^{N_t}$ the array response and steering  vectors evaluated at elevation $\theta_{\ell}^r$ ($\theta_{\ell}^t$) and azimuth $\phi_{\ell}^r$ ($\phi_{\ell}^t$) arrival (departure) angles at the receiver (transmitter), respectively. Each channel matrix realization is normalized such that $\mbb{E} \left[ \frob{\mb{H}}^2 \right] = N_r N_t$. We assume that each path has the same average power, i.e., $\alpha_{\ell}$ is modeled as circular symmetric Gaussian random variables with zero mean and unit variance. The departure/arrival elevation and azimuth angles are uniformly distributed in the intervals $\openintervalboth{-\pi/2}{\pi/2}\,\si{\radian}$ and $\closedinterval{0}{2\pi}\,\si{\radian}$, respectively.
	
	The transmit and receive arrays are assumed to be linear and uniformly spaced with inter-element spacing of $\lambda/2$, where $\lambda$ denotes the carrier wavelength. The methods discussed in this paper can be easily applied to arrays with arbitrary geometry and element beam-pattern. Assuming that the uniform linear array is distributed along the $x$-axis, the steering and array response vectors follow the Vandermonde structure
	\begin{equation}
		\mb{a}_x(\theta, \phi) = \frac{1}{\sqrt{N_x}}\left[ 1, \ldots, e^{j\pi(N_x-1)\sin\theta\,\cos\phi} \right]^\tran \in \mbb{C}^{N_x},
	\end{equation}
	for $x \in \{t, r\}$.
	
	\subsection{Power Consumption and Loss Models} \label{sec:pow}

	Energy efficiency is an important concern in the design of mmWave massive MIMO systems. Large antenna arrays are employed to compensate for the increased path loss of mmWave channels and to achieve high spectral efficiency. However, several electronic components such as DACs, phase-shifters, and power amplifiers become inefficient when operating in the millimeter wave range over large bandwidth, making radio systems equipped with such arrays expensive due to large power consumption and losses of their RF front-end. Therefore, there has been important research efforts on designing systems that not only maximize their spectral efficiency, but also their energy efficiency, which can be defined as the ratio of spectral efficiency to power consumption~\cite{abbas_millimeter_2016}. 
	
	Hybrid systems seek to decrease power consumption by employing only a few RF chains and an analog beamforming stage. Although hybrid architectures are expected to exhibit reduced power consumption compared to fully-digital systems, they shall also present important power losses due to phase-shifters, power dividers, and power combiners. To compensate for these losses, one could simply adopt high-gain PAs that would cover the insertion losses. However, high-gain and high-efficiency PAs are not available for mmWave yet, and, thus, these losses cannot be easily compensated for, incurring in energy efficiency degradation. In order to consider these power losses in our modeling, we adopt a realistic RF formulation based on the insertion loss of power dividers, power combiners, and phase-shifters.
	
	Recently, active and passive phase-shifting elements for mmWave have been reported in \cite{huang_60_2017}. The former present non-negligible power consumption and moderate gain, whereas the latter has insignificant power consumption and considerable insertion loss. Therefore, the system designer can either maximize the transmit power with increased power consumption by active phase-shifters, or reduce power consumption with decreased transmit power by passive phase-shifters. Which circuit implementation leads to the most energy-efficient hybrid system is not clear. Thus, in order to compare the energy performance of the different transmitter architectures, we define power consumption models, and we introduce the power loss modeling in the following.
	
	The influence of signal processing on system power consumption may be important in massive MIMO systems. It is well-known that the computational demand of standard signal processing methods significantly grows with the number of antenna elements. Since these methods are typically implemented in power hungry devices, it is thoughtful to avoid extensive computations in order to save energy. As pointed out in \cite{prasad_energy_2017}, this aspect is usually overlooked and should be considered during system optimization. Therefore, in our modeling, we consider the \emph{computational} power consumption corresponding to precoder optimization, and the \emph{static} power consumption of the RF front-end.
	
	Let us first discuss the static power consumption of hybrid and digital transmitters. The transmitter RF front-end is formed by a DAC for each I/Q channel, RF chains, PSN (hybrid systems), and PAs. The considered direct conversion RF chain consists of two low-pass filters, two mixers, a local oscillator shared among all chains, and a $90^\circ$ hybrid with buffers. Denoting the power consumption of low-pass filter, mixer, local oscillator, hybrid with buffer as $P_\text{LP}$, $P_\text{M}$, $P_\text{LO}$, $P_\text{H}$, respectively, the power consumption $\Prf$ of a single RF chain is then 
	\begin{equation} \label{eq:rfchain}
	\Prf = 2P_{\text{LP}} + 2 P_\text{M} + P_\text{H}.
	\end{equation}
	A transmitter with digital precoding employs $N_t$ RF chains, PAs, and pairs of DACs, thus its power consumption is given~by
	\begin{equation}
	P_\text{D}(N_t,b_\text{DAC},F_s) = P_{\text{LO}}  + P_{\text{PA}} + N_t[2P_{\text{DAC}}(b_\text{DAC},F_s) + \Prf],
	\end{equation}
	where $P_{\text{DAC}}(b_\text{DAC},F_s)$ denotes the power consumption of a DAC sampling at $F_s$ Hertz with $b_\text{DAC}$ bits of resolution. $P_\text{PA}$ stands for the power consumed by all PAs. A hybrid transmitter with fully-connected PSN, by contrast, employs $L_t$ DAC/RF chain pairs. A total of $N_t L_t$ phase-shifters are used in this architecture, thus its static power consumption is
	\begin{align}
		&P_\text{FPSN}(N_t,L_t,b_\text{DAC}, b_\text{PS}, F_s) = \\
		&P_{\text{LO}} + P_{\text{PA}} + L_t[2P_{\text{DAC}}(b_\text{DAC}, F_s) + \Prf ] +  N_t  L_t P_{\text{PS}}(b_\text{PS}),
	\end{align}
	where $P_{\text{PS}}(b_\text{PS})$ denotes power consumption of a single phase-shift element with $b_\text{PS}$ bits of resolution. In general, the power consumption of dividers and combiners is negligible. Transmitters with partially-connected PSN have $N_a$ phase-shifters per sub-array, i.e., there are $N_a L_t$ phase-shifters in total, hence their static power consumption is given by 
	\begin{align}
	&P_\text{PPSN}(N_t,L_t,b_\text{DAC}, b_\text{PS}, F_s) = \\
	&P_{\text{LO}} +P_{\text{PA}} + L_t[2P_{\text{DAC}}(b_\text{DAC}, F_s) + \Prf ] +  N_a L_t P_{\text{PS}}( b_\text{PS}).	
	\end{align}
	
	The need of efficient mmWave systems fueled the development of new electronic components. One can find in the literature different parameters for the electronic devices considered in our models. We adopt an optimistic parameter selection approach, in which we choose the most efficient implementation reported. Results obtained with this approach ought to provide an idea of what to expect for future mmWave massive MIMO systems. Regarding the power consumption of the RF chain components, we consider values reported in \cite[Chapter 5]{marcu_lo_2011} for the $90^\circ$ hybrid with buffers, \cite{jin_7_2011} for the mixers, \cite{rangan_energy_2013} for the low-pass filters, and \cite{scheir_52_2008} for the local oscillator. We employ DACs with a binary-weighted current-steering topology. Its power consumption is a function of the effective number of bits $b_\text{DAC}$ and the sampling frequency $F_s$~\cite{cui_energy-constrained_2005}:
	\begin{equation} \label{eq:pdac}
	P_\text{DAC}(b_\text{DAC},F_s) = 1.5 \times 10^{-5} \cdot 2^{b_\text{DAC}} + 9\times10^{-12} \cdot b_\text{DAC} \cdot F_s\,.
	\end{equation}
	Equation~\eqref{eq:pdac} was obtained by the same parameter setup as in~\cite{cui_energy-constrained_2005}. According to \cite[Section 1]{olieman_time-interleaved_2016}, this type of DAC is well-adapted for high-speed conversion since no buffer is required and thus switching can be done fast. PAs are the most power hungry devices on the transmit side, because the high linearity requirements render them inefficient. The power consumed by the set of PAs with power-added efficiency (PAE) $\eta$ is given by $P_\text{PA} = P_x/\eta$ \cite{gao_power-performance_2017}, where $P_x$ is the actual transmit power considering RF losses defined in Section~\ref{sec:precstr}. Peak PAE values for state-of-art PAs are listed in \cite{greene_60-ghz_2017}, and vary between $6.5$$-$$27\%$. According to our optimistic approach, we set $\eta = 27\%$. State-of-art mmWave active phase-shifters have been listed in \cite{mendez-rial_hybrid_2016} and their power consumption lies in the range $15$$-$$108\,\si{\milli\watt}$ with moderate peak gains. As an alternative to active phase-shifters, passive implementations with negligible power consumption and significant insertion loss have been reported in \cite{huang_60_2017}. We spend $P_\text{PS} = 21.6\,\si{\milli\watt}$ for active phase-shifters \cite{pepe_78.8-92.8_2015} and consider zero power consumption for passive phase-shifters \cite{huang_60_2017}. The assumed power consumption values of the RF front-end components are summarized in Table~\ref{tab:pc}.
	
	Let us now present the power loss model considered in this work. As in \cite{garcia-rodriguez_hybrid_2016}, RF power losses in the analog beamformer lead to the following representation of the analog precoding matrix: $\Frf/\sqrt{\Lrf^{\{\text{FPSN},\text{PPSN}\}}}$, where $\Lrf^{\{\text{FPSN},\text{PPSN}\}}$ denotes the loss factor for fully- and partially-connected PSNs. We will derive expressions for $\Lrf^{\{\text{FPSN},\text{PPSN}\}}$ taking into account the insertion loss of each RF component in the PSN. We consider that $K$-way devices are formed by concatenating $\lceil \log_2(K) \rceil$ stages of two-way devices~\cite{garcia-rodriguez_hybrid_2016}, thus the power loss for the total dividing and combining stages is given by $L_{\{\text{D},\text{C}\}}(K)\,[\si{\decibel}] = \bar{L}_{\{\text{D},\text{C}\}} \lceil \log_2(K) \rceil$, where $\bar{L}_{\{\text{D},\text{C}\}}$ denotes the static power loss in decibels of a single two-way device~\cite{garcia-rodriguez_hybrid_2016}. We consider that two-way power dividers lose $0.6\,\si{\decibel}$~\cite{garcia-rodriguez_hybrid_2016}, and two-way power combiners $0.6\,\si{\decibel}$ plus extra $3\,\si{\decibel}$ due to amplitude and phase mismatches. Regarding phase-shifting, we assume that active phase-shifters have a gain of $2.3\,\si{\decibel}$, i.e., loss of $L_\text{PS} = -2.3\,\si{\decibel}$~\cite{pepe_78.8-92.8_2015}, while passive phase-shifters have a loss of $L_\text{PS}=8.8\,\si{\decibel}$~\cite{huang_60_2017}. Values for $\bar{L}_\text{D}$, $\bar{L}_\text{C}$, and $L_\text{PS}$ are listed in Table~\ref{tab:pc}.
	
	Let us consider a fully-connected PSN fed with $P_{\text{in}}$ Watts. The signal power at the input of each $N_t$-way power divider is $P_{\text{in}}/L_t$. After power splitting and phase-shifting, the signal power at each network branch is $(P_{\text{in}}/L_t)/(N_t   L_\text{D}(N_t)   L_\text{PS})$. The power level at the output of each $L_t$-way power combiner is $[(P_{\text{in}}/L_t)  L_t]/(N_t  L_\text{D}(N_t)  L_\text{PS}  L_\text{C}(L_t) )$. Since there are $N_t$ power combiners, the PSN output power is $P_{\text{out}} =P_{\text{in}}/ ( L_\text{D}(N_t)  L_\text{PS}  L_\text{C}(L_t) )$, and, hence, the loss factor for fully-connected PSNs is
	\begin{equation}
	\Lrf^{\text{FPSN}} = \frac{P_{\text{in}}}{P_\text{out}} = L_\text{D}(N_t)   L_\text{PS}   L_\text{C}(L_t).
	\end{equation}
	The loss factor for partially-connected PSNs is obtained analogously, considering $N_a$-way power dividers and no power combining:
	\begin{equation}
	\Lrf^{\text{PPSN}} = L_\text{D}(N_a)  L_\text{PS}.
	\end{equation}
	
	\begin{table}[!t]
		\renewcommand{\arraystretch}{1.3}
		\caption{Power consumption and loss of the RF front-end components.}
		\label{tab:pc}
		\centering
		\begin{tabular}{c|c|c}
			\bfseries Component& \bfseries Notation & \bfseries Value\\
			\hline
			Power amplifier \cite{greene_60-ghz_2017} & $P_\text{PA}$ 	   & $P_x/\eta,\,\eta=27\%$ \\
			Phase-shifter (active \cite{pepe_78.8-92.8_2015} $;$ passive \cite{huang_60_2017}) & $P_\text{PS}$ 	    & $21.6\,;\,0\,\si{\milli\watt}$ \\
			DAC~\cite{cui_energy-constrained_2005}	& $P_\text{DAC}$      & Equation~\eqref{eq:pdac}\\ 
			Local oscillator \cite{scheir_52_2008} & $P_\text{LO}$        & $22.5\,\si{\milli\watt}$			\\
			$90^\circ$ hybrid with buffers \cite{marcu_lo_2011}	& $P_\text{H}$ 	     & $3\,\si{\milli\watt}$\\
			Mixer \cite{jin_7_2011}	& $P_\text{M}$ 	    & $0.3\,\si{\milli\watt}$\\
			Low-pass filter \cite{rangan_energy_2013}	& $P_\text{LP}$        & $14\,\si{\milli\watt}$\\
			RF chain (Equation~\eqref{eq:rfchain}) & $\Prf$ 	& $31.6\,\si{\milli\watt}$\\
			\hline
			Two-way power divider \cite{garcia-rodriguez_hybrid_2016} 	& $\bar{L}_\text{D} $ 	   & $0.6\,\si{\decibel}$ \\
			Two-way power combiner \cite{garcia-rodriguez_hybrid_2016} 	& $\bar{L}_\text{C} $ 	   & $0.6\,\si{\decibel} + 3\,\si{\decibel}$ \\
			Phase-shifter (active \cite{pepe_78.8-92.8_2015} $;$ passive \cite{huang_60_2017}) 	& $L_\text{PS} $ 	   & $-2.3$ $;$ $8.8 \,\si{\decibel}$			
		\end{tabular}
	\end{table}
	
	The computational analysis of signal processing methods involves counting of arithmetical operations, memory overhead analysis, among other factors~\cite{golub_matrix_2012}. Here, we consider the simple computational power consumption model of \cite{bjornson_optimal_2015}, in which the power spent with precoder optimization is proportional to the number $N_\text{flops}$ of floating-point operations (flops) demanded by the optimization algorithm, the number $C$ of coherence blocks per second, and inversely proportional to the transmitter computational efficiency $E_c$ [flops/s/W], i.e., it is given by 
	\begin{equation} 
	P_\text{comp} = C N_\text{flops}/E_c. 
	\end{equation}
	The number $C$ of coherence blocks determines how many times the system updates its precoding filters per second. It is defined as the ratio of the transmit bandwidth $B$ to the number of coherence blocks, which is given by the product of the channel coherence bandwidth $B_C$ and the channel coherence time $T_C$, i.e., $C=B/(B_C T_C)$~\cite{bjornson_optimal_2015}. Naturally, the computational power consumption in practice strongly depends on the hardware implementation. However, distinct implementations are characterized by different computational efficiency $E_c$. Therefore, the considered model is general enough to provide insights on the computational power consumption at the transmitter.

	\section{Precoding Strategies} \label{sec:precstr}
	
	In this section, we introduce the quantized hybrid precoding problem, and present precoding strategies to solve it. Analog precoding methods for fully- and partially-connected PSN are presented in Sec.~\ref{sec:analprec}, baseband precoding is defined in Sec.~\ref{sec:basebandprec}, computational complexity analysis is conducted in Sec.~\ref{sec:companalysis}, and achievable rate lower bounds are derived in Sec.~\ref{sec:achrate}.
	
	The \emph{quantized hybrid precoding problem} is based on system model \eqref{eq:eqmimo_white} assuming perfect and instantaneous channel state information at both transmitter and receiver. It consists of finding the precoding matrices $\Frf$ and $\Fbb$ that maximize the achievable instantaneous rate $R$~\cite{biglieri_mimo_2007}, i.e.,
	
	\begin{small}
		\begin{maxi}|l| 
			{\Frf, \Fbb}{\log_2 \det \left( \mb{I}_{N_r} +  \tfrac{1}{\Lrf}\mb{R}_{n_G n_G}^{-1/2} \mb{H}'\mb{R}_{uu} \mb{H}^{'\hermit} \mb{R}_{n_G n_G}^{-1/2,\hermit} \right) }{\label{prob:hp}}{}
			\addConstraint{[\Frf]_{u,v} \in \sFrf,\,\forall u \forall v,\,\mbb{E}\left[ \|\tilde{\mb{x}} \|_2^2 \right] \leq \Pmax}.
		\end{maxi}
	\end{small}

	The achievable rate expression in \eqref{prob:hp} provides some insights on the performance of the quantized hybrid precoder. In order to discuss them, consider the actual transmited power $P_x = \esp{\|\mb{x}\|_2^2 } = \frac{1}{\Lrf} P_{\tilde{x}}$, where $P_{\tilde{x}} = \esp{\|\tilde{\mb{x}}\|_2^2 }  = (1-\rho_b) \frob{\Frf \Fbb}^2 + \tr(\Frf \mb{R}_{ee}\Frf^\hermit)$ stands for the lossless transmit signal power. In the appendix, we show that $P_{\tilde{x}} = \Pmax$ when the baseband precoder presented in Section~\ref{sec:basebandprec} is employed. Therefore, the actual transmit power is $P_x = \Pmax/\Lrf$, satisfying the average power constraint. As discussed in Section~\ref{sec:pow}, if the RF losses are not properly compensated for, then the actual transmit power is reduced, decreasing the achievable rate.

	Problem~\eqref{prob:hp} is sufficiently general to model other precoding schemes. For example, when $\rho_b=0$, we have $\mb{R}_{ee} = \mb{0}_{M\times M} \Rightarrow \mb{R}_{n_G n_G}^{-1/2} = \sigma_n^{-1} \mb{I}_{N_r}$, and the objective function is rewritten as $\log_2 \det \left( \mb{I}_{N_r} +  \tfrac{1}{\Lrf}\mb{R}_{n_G n_G}^{-1} \Heq \mb{R}_{uu} \Heq^\hermit \right)$, which refers to the hybrid precoding problem \cite{el_ayach_spatially_2014} with lossy RF hardware. Moreover, when $\Fbb \in \mbb{C}^{N_t \times N_s}$, $\Frf = \mb{I}_{N_t}$, $\Lrf = 1$, and $\rho_b = 0$, \eqref{prob:hp} falls back to the classical SU-MIMO digital precoding problem. The corresponding quantized problems are obtained for $\rho_b \neq 0$.
	
	To solve \eqref{prob:hp}, we adopt the sub-optimal strategy of decoupling the precoder design problem into two subproblems for optimizing $\Frf$ and $\Fbb$, separately. We first tackle the analog precoding subproblem for obtaining $\Frf$, allowing us to form the equivalent channel matrix $\Heq$. Subsequently, we design our baseband precoder based on the SVD of $\Heq$.
	
	\subsection{Analog Precoding Strategies} \label{sec:analprec}
	
	In the following, we present analog precoding methods for the fully- and partially-connected PSN topologies. Since $\Frf$ may have very large dimensions, the computational efforts of precoding design can be significant, and, thus, we focus on low computational complexity instead of high spectral efficiency.
	
	\subsubsection{Fully-connected PSN} \label{sec:fpsn}
	
	As in \cite{abbas_millimeter_2016}, we design the RF precoder with the alternating projection method of \cite[Section 3]{tropp_designing_2005}. It consists of initially selecting the first $L_t$ right singular vectors of the MIMO channel matrix $\mb{H}$ and forming a semi-unitary precoder $\mb{F}_{\text{SU}}$. Next, this matrix is projected onto the constant modulus space, and the result is projected back to the semi-unitary matrix $\mb{F}_{\text{SU}}$. This alternating projection procedure is repeated until convergence, which is achieved when the Frobenius norm residual between two consecutive iterations is smaller than a tolerance value. After convergence, the phase of each element in $\Frf$ is quantized to the closest value in the phase resolution set $\mc{F}_\text{RF}$. The fully-connected PSN hybrid precoder is summarized in Algorithm~\ref{alg:fpsn}.
	
	\begin{algorithm}
		\caption{Hybrid precoding (fully-connected PSN)}
		\label{alg:fpsn}
		\begin{algorithmic}[1]
			\Require{$\mb{H}$, $L_t$, $b_\text{PS}$}
			\State Compute SVD of $\mb{H} = \mb{U} \mb{\Sigma} \mb{V}^\hermit$ \label{line:am_svd_full}
			\State Initialize $\mb{F}_\text{SU} \gets [\mb{V}]_{:,1:L_t}$
			\Repeat
			\State $\Frf \gets\exp(j\angle \mb{F}_\text{SU})$ \label{line:am_norm}
			\State Compute SVD of $\Frf = \tilde{\mb{U}} \tilde{\mb{\Sigma}} \tilde{\mb{V}}^\hermit$	\label{line:am_svd}
			\State $\mb{F}_\text{SU} \gets [\tilde{\mb{U}}]_{:,1:L_t} \tilde{\mb{V}}^\hermit$	\label{line:am_projback}
			\Until{convergence criterion triggers}
			\State $\Frf \gets \exp(j \mc{Q}_{b_\text{PS}}( \angle \Frf))$
		\end{algorithmic}
	\end{algorithm}
	
	\subsubsection{Partially-connected PSN} \label{sec:ppsn}
	
	For the partially-connected RF precoding, we define the sub-channel matrix $\mb{H}_\ell = [\mb{H}]_{:,n_a + (\ell-1)N_a} \in \mbb{C}^{N_r \times N_a},\,n_a = 1,\ldots,N_a$, that contains the $N_a$ columns of $\mb{H}$ belonging to the $\ell$-th antenna sub-array. The proposed design employs quantized MET for each sub-array, i.e., $\mb{f}_\ell = \exp(j \mc{Q}_{b_\text{PS}}(\angle \mb{v}_\ell^{\text{max}})), \,\ell=1,\ldots,L_t$, where $\mb{v}_\ell^{\text{max}}$ denotes the right singular vector corresponding to the largest (dominant) singular value of $\mb{H}_\ell$. The projection step enforces the constant modulus constraint imposed by the phase-shifters. Since we are interested only in $\mb{v}_\ell^{\text{max}}$, we can use the simple power method \cite[Section 8.2.1]{golub_matrix_2012} to calculate it. This algorithm is significantly less expensive than, for example, the $R$-SVD algorithm~\cite{golub_matrix_2012} which is useful for computing the full SVD. The power method achieves convergence when the residual error between two consecutive iterations is smaller than a tolerance value. It is guaranteed to converge if the dominant singular value is larger than all the other singular values in modulus and if the initial guess for $\mb{v}_\ell^{\text{max}}$ has a non-zero component in the direction of the corresponding right singular vector \cite{golub_matrix_2012}, which occurs for the model presented in Section~\ref{sec:channel}. The partially-connected hybrid precoding method is outlined in Algorithm~\ref{alg:ppsn}.
	
	\begin{algorithm}
		\caption{Analog beamforming for partially-connected PSN via the power method}
		\label{alg:ppsn}
		\begin{algorithmic}[1]
			\Require{ $\mb{H}$, $L_t$, $N_a$, $b_\text{PS}$}
			\For{$\ell = 1,\ldots, L_t$}
				\State Form $\mb{H}_\ell \gets [\mb{H}]_{:,n_a + (\ell-1)N_a},\, n_a=1,\ldots,N_a$
				\State Randomly initialize $\mb{v}_\ell \in \mbb{C}^{N_a}$
				\Repeat
					\State $\mb{v}_\ell \gets \mb{H}_\ell^\hermit \mb{H}_\ell \mb{v}$ \label{line:power}
					\State $\mb{v}_\ell \gets \mb{v}_\ell/\|\mb{v}_\ell\|_2$ \label{line:norm}		
				\Until{convergence criterion triggers}
			\EndFor
			\State $\Frf \gets \Diagblk\left[ \exp(j \mc{Q}_{b_\text{PS}}(\angle \mb{v}_1)), \ldots, \exp(j \mc{Q}_{b_\text{PS}}(\angle \mb{v}_{L_t})) \right]$ \label{line:project}
		\end{algorithmic}
	\end{algorithm}
	
	\subsection{Digital Precoding Strategy} \label{sec:basebandprec}
	
	Unfortunately, finding the digital precoder $\Fbb$ that maximizes \eqref{prob:hp} for fixed $\Frf$ is not straightforward. The SVD precoding with water-filling power allocation is not guaranteed anymore to decompose the MIMO system into orthogonal sub-channels due to the structure of the total noise vector $\mb{n}'$. Furthermore, in order to form $\mb{H}'$, one needs knowledge of the diagonal elements of $\mb{R}_{uu}$, which depends on $\Fbb$. 
	
	To solve these problems, notice that as the quantization resolution increases, $\rho_b \rightarrow 0$, $\mb{R}_{n_G n_G}  \rightarrow \mb{R}_{nn}$, $\mb{H}' \rightarrow \Heq$, and the instantaneous rate in \eqref{prob:hp} goes to $\log_2 \det\left( \mb{I}_{N_r} + \frac{1}{\Lrf}\mb{R}_{n n}^{-1}\Heq\mb{R}_{uu}\Heq^\hermit \right)$. In this case, SVD precoding and water-filling power allocation with respect to $\Heq$ and $\mb{R}_{nn}$ become optimal. Note that this strategy still maximizes the achievable rate regardless of $\frac{1}{\Lrf}$, thus we ignore this power loss factor in the transmitted signal model when designing $\Fbb$ by setting $\Lrf =1$. In general, this strategy becomes sub-optimal under low-resolution quantization. Nevertheless, it is still expected to be close to optimal, particularly at low SNR. In such regime, AWGN overruns the quantization noise, motivating the use of the SVD precoding. At high SNR, however, the quantization noise dominates and this solution is only sub-optimal. Since massive MIMO systems will likely operate in low SNR regimes, this choice is reasonable in practice. 
	
	Following our discussion, we design $\Fbb$ as the optimal precoder in infinite-resolution DAC scenarios. In this sense, we assume $\rho_b=0$ and knowledge of $\Frf$ (see Section~\ref{sec:analprec}), allowing $\Heq$ to be formed\footnote{In a practical setup, $\Heq$ could be estimated at the receiver and sent back to the transmitter via a feedback channel.}. The baseband precoder is then obtained by solving
	\begin{maxi}|l| 
		{\Fbb}{\log_2 \det \left( \mb{I}_{N_r} + \mb{R}_{nn}^{-1} \Heq \Fbb \Fbb^\hermit \Heq^{\hermit} \right)}{\label{prob:hp_bb}}{}
		\addConstraint{\mbb{E}\left[ \|\tilde{\mb{x}} \|_2^2 \right]  \leq \Pmax}.
	\end{maxi}
	It is straightforward to show that the average power constraint in \eqref{prob:hp_bb} can be rewritten as $\frob{ \Frf \Fbb }^2 \leq \Pmax$ when $\rho_b=0$. Let the SVD of the equivalent channel be denoted by $\Heq = \mb{U} \mb{\Sigma} \mb{V}^\hermit$, in which $\mb{U} \in \mbb{C}^{N_r \times N_s}$, $\mb{\Sigma} \in \mbb{C}^{N_s \times N_s}$, and $\mb{V} \in \mbb{C}^{M \times N_s}$. We employ SVD baseband precoding with water-filling power allocation, and normalize it so that the average power constraint in \eqref{prob:hp_bb} is obeyed: 
	\begin{equation} \label{eq:baseband}
	\Fbb = \tfrac{\sqrt{\Pmax}}{\frob{\Frf \mb{Q}}} \mb{Q},
	\end{equation}
	where $\mb{Q} = \mb{V} \wfsqrt \in \mbb{C}^{N_t \times N_s}$ represents the SVD precoder with the diagonal power allocation matrix $\wf \in \mbb{R}^{N_s \times N_s}$. In the appendix, we show that the average power constraint is always satisfied regardless of DAC quantization resolution, and that $\frob{\Frf \mb{Q}}^2=N_t\Pmax$, which gives $\Fbb = \frac{1}{\sqrt{N_t}} \mb{Q}$.
	
	For fully-digital transmitters, Problem~\eqref{prob:hp_bb} becomes
	\begin{maxi}|l| 
		{\Fbb}{\log_2 \det \left( \mb{I}_{N_r} + \mb{R}_{nn}^{-1} \mb{H} \Fbb \Fbb^\hermit  \mb{H}^\hermit  \right)}{\label{prob:dig}}{}
		\addConstraint{\frob{\Fbb}^2 \leq \Pmax}.
	\end{maxi}
	The solution in this case is $\Fbb = \bar{\mb{V}}\bar{\mb{\Lambda}}^{-1/2}$, where $\bar{\mb{V}} \in \mbb{C}^{N_t \times N_s}$ represents the right singular vector matrix of $\mb{H}$, and $\bar{\mb{\Lambda}} \in \mbb{C}^{N_s \times N_s}$ the corresponding water-filling power allocation matrix.
	
	\subsection{Computational Complexity Analysis} \label{sec:companalysis}
	
	The computational analysis of the proposed precoding methods is important to evaluate the impact of signal processing on the transmitter power budget with the power consumption modeling presented in Section~\ref{sec:pow}. In the following analysis, we consider that the matrix product of $\mb{A} \in \mbb{C}^{M \times R}$ and $\mb{B} \in \mbb{C}^{R\times N}$ requires $2MNR$ flops as in \cite{golub_matrix_2012}.

	From our discussion in Section~\ref{sec:basebandprec}, fully-digital precoding consists of SVD and water-filling power allocation. Its computational complexity is dominated by the algorithm that decomposes $\mb{H}$. According to \cite[Figure 8.6.1]{golub_matrix_2012}, one needs $N_\text{flops}^{\text{SVD}} = 4P^2Q + 22Q^3$ flops to compute the full SVD of a $(P \times Q)$-dimensional matrix with the $R$-SVD algorithm. Therefore, the number of flops necessary to obtain the digital precoder is $N_\text{flops}^{\text{D}} = 4N_r^2 N_t + 22 N_t^3$.
	
	Signal processing of hybrid precoding is carried out in two stages: baseband and analog precoding. The computational complexity of the former stage is dominated by the SVD of $\Heq$ which demands $N_\text{flops}^{\text{SVD}} = 4N_r^2 L_t + 22L_t^3 + 2N_r N_t L_t$ flops. The term $2N_r N_t L_t$ in $N_\text{flops}^{\text{SVD}}$ refers to the matrix product that forms $\Heq$. The computational effort of the latter stage corresponds to the number of flops demanded by the analog precoding algorithm. The flop counting for Algorithms \ref{alg:fpsn} and \ref{alg:ppsn} is detailed in Tables~\ref{tab:fpsn} and \ref{tab:ppsn}, respectively. According to Table~\ref{tab:fpsn}, $N_\text{flops}^{\text{FPSN}} = 4N_r^2 N_t + 22 N_t^3 + I(4N_t L_t + 4N_t^2 L_t + 22L_t^3 + 2N_t L_t^2 )$ flops are necessary to compute the fully-connected analog precoder, where $I$ denotes the number of iterations of the alternating minimization method. The total cost is then given by the baseband and analog precoders computation cost $N_\text{flops}^{\text{F}} = N_\text{flops}^{\text{SVD}} + N_\text{flops}^{\text{FPSN}}$. According to Table~\ref{tab:ppsn}, the total number of flops necessary for computing the partially-connected analog precoder is $N_\text{flops}^{\text{PPSN}} = J(4N_t N_a + 2N_a + 1)$, where $J$ represents the number of iterations of the power method. The total computation cost of the partially-connected hybrid precoder accounts for the calculation of $\Heq$ and its SVD. Therefore, it is given by $N_\text{flops}^{\text{P}} = N_\text{flops}^{\text{SVD}} + N_\text{flops}^{\text{PPSN}}$. Comparing $N_\text{flops}^{\text{F}}$ and $N_\text{flops}^{\text{P}}$, we observe that the latter is less complex than the former because the partially-connected structure is exploited to reduce the analog precoding computational complexity.	
	
		\begin{table}[!t]
		\renewcommand{\arraystretch}{1.3}
		\caption{Flop counting for Algorithm~\ref{alg:fpsn}.}
		\label{tab:fpsn}
		\centering
		\begin{tabular}{c|c}
			\bfseries Line & \bfseries Flops count\\
			\hline
			\ref{line:am_svd_full}	  & $4N_r^2 N_t + 22 N_t^3$\\
			\ref{line:am_norm} 		 & $4N_t L_t$ \\
			\ref{line:am_svd}  		   & $4N_t^2 L_t + 22L_t^3$\\
			\ref{line:am_projback}	& $2N_t L_t^2 $ \\
		\end{tabular}
	\end{table}

	\begin{table}[!t]
		\renewcommand{\arraystretch}{1.3}
		\caption{Flop counting for Algorithm~\ref{alg:ppsn}.}
		\label{tab:ppsn}
		\centering
		\begin{tabular}{c|c}
			\bfseries Line & \bfseries Flops count\\
			\hline
			\ref{line:power}		& $4N_t N_a$ \\
			\ref{line:norm}			& $2N_a + 1$\\
		\end{tabular}
	\end{table}	
	
	\subsection{Achievable Rate Bounds} \label{sec:achrate}
	
	Now we conduct an asymptotic performance assessment of the quantized hybrid precoder. Let us first consider the received signal model \eqref{eq:approx_signal_ftw} with Gaussian noise approximation and hybrid precoding, then we employ SVD combining (with respect to $\Heq = \mb{U} \mb{\Sigma} \mb{V}^\hermit$) at the receiver, and finally decorrelate the Gaussian noise vector $\mb{n}_G$ to obtain an achievable rate lower bound. From Equation~\eqref{eq:baseband}, it follows that the rotated received signal can be written as 

	\begin{small}
		\begin{align}
			\tilde{\mb{y}} &= \tU^\hermit \mb{y} = \tfrac{1}{\sqrt{\Lrf N_t}}\tU^\hermit \Heq \Ups \mb{V} \wfsqrt \mb{s} + \tfrac{1}{\sqrt{\Lrf}}\tU^\hermit \Heq \mb{e} + \tU^\hermit \mb{n}\\
						   &= \sqrt{\tfrac{1-\rho_b}{\Lrf N_t}} \tS \wfsqrt \mb{s} + \tilde{\mb{n}} \in \mbb{C}^{N_s}, 		\label{eq:roteqmimo}
		\end{align}
	\end{small} 

	where $\tilde{\mb{n}} = \frac{1}{\sqrt{\Lrf}}\tS \tV^\hermit \mb{e} + \tU^\hermit  \mb{n} \in \mbb{C}^{N_s}$ represents the rotated total additive Gaussian noise component with covariance matrix $\mb{R}_{\tilde{n} \tilde{n}} = \frac{1}{\Lrf}\tS \tV^\hermit \mb{R}_{ee} \tV \tS + \sigma_n^2 \mb{I}_{N_s}$. After the whitening filter, we have
	\begin{equation} \label{eq:roteqmimo_white}
		\mb{R}_{\tilde{n} \tilde{n}}^{-1/2} \tilde{ \mb{y} } = \sqrt{\tfrac{1-\rho_b}{\Lrf N_t}}  \mb{R}_{\tilde{n} \tilde{n}}^{-1/2}  \tS \wfsqrt \mb{s} +  \mb{R}_{\tilde{n} \tilde{n}}^{-1/2}\tilde{\mb{n}}.
	\end{equation}
	This equation provides the following lower bound for the achievable rate:
	\begin{align} \label{eq:lbachrate}
	R \geq \log_2 \det \left( \mb{I}_{N_s} + \tfrac{1-\rho_b}{\Lrf N_t}  \mb{R}_{\tilde{n} \tilde{n}}^{-1/2} \tS \wf \tS \mb{R}_{\tilde{n} \tilde{n}}^{-1/2,\hermit}\right).
	\end{align}
	For low-resolution quantization and high SNR, water-filling power allocation yields $\wf = (\Pmax/N_s) \mb{I}_{N_s}$ and \eqref{eq:lbachrate} goes to
	\begin{equation} \label{eq:lbachrate_lowRES_highSNR}
	R \geq \log_2 \det \left( \mb{I}_{N_s} + \tfrac{\Pmax(1-\rho_b)}{\Lrf N_t N_s}  \mb{R}_{\text{high}}^{-1/2} \tS^2 \mb{R}_{\text{high}}^{-1/2,\hermit} \right),
	\end{equation}
	where $\mb{R}_{\text{high}} = \tS \tV^\hermit \mb{R}_{ee} \tV \tS$. At low SNR, $\mb{R}_{\tilde{n} \tilde{n}} \rightarrow \sigma_n^2 \mb{I}_{N_s}$, and the achievable rate is lower bounded by $R \geq \log_2 \left[ 1 + (1-\rho_b)\frac{\gamma}{\Lrf N_t}\sigma_\text{max}^2\right]$, where $\sigma_\text{max}$ denotes the largest singular value of $\Heq$ and $\gamma$ the SNR. Inequality \eqref{eq:lbachrate_lowRES_highSNR} shows that quantization noise dominates over AWGN in high SNR regime, leading to saturation of the system capacity. The obtained achievable rate expression at low SNR shows that quantized system capacity is always smaller than that of unquantized systems due to the factor $(1-\rho_b)$ inside the logarithm.		

	\section{Simulation Results} \label{sec:sim}
	
	In this section, we present the results of numerical simulations conducted to evaluate the spectral and energy efficiencies of the proposed precoding strategies. Also, the power consumption model introduced in Section~\ref{sec:pow} is assessed. Hereafter, hybrid precoding with fully- and partially-connected PSN are referred to as HPF and HPP, respectively. The obtained results are averaged over $1000$ independent experiments. For each experiment realization, a channel matrix with $L=5$ paths is generated according to Equation~\eqref{eq:channel_vector}. We assume that both transmitter and receiver have perfect and instantaneous channel state information. We set the average transmit power constraint to $\Pmax = 1\si{\watt}$, which is reasonable for base stations with small coverage. Also, in order to reduce costs, the number $L_t$ of RF chains is equal to the number $N_s$ of streams. We assume phase-shifters with phase range of $360^\circ$ and shift resolution of  $b_\text{PS} = 5$ bits, as in the hardware implementation of \cite{huang_60_2017}. The DAC sampling rate is set to $F_s = 1 \,\si{\giga \hertz}$, which should be sufficient for mmWave systems to provide high data rates. Regarding the calculation of computational power consumption, we consider the same parameters as in~\cite{bjornson_optimal_2015}: computational efficiency of $E_c=12.8$ Gflops/s/$\si{\watt}$, transmission bandwidth of $B=20\,\si{\mega\hertz}$, channel coherence bandwidth of $B_C = 180\,\si{\kilo\hertz}$, and channel coherence time of $T_C = 10\,\si{\milli\second}$, giving $C=B/(B_C T_C) = 11111$ coherence blocks per second. The convergence tolerance value of the iterative algorithms is set to $10^{-6}$. Preliminary simulations have shown that the iterative algorithms used to compute HPF and HPP converge within $37$ and $11$ iterations on average, respectively.
	
	\subsection{Spectral Efficiency} 
	
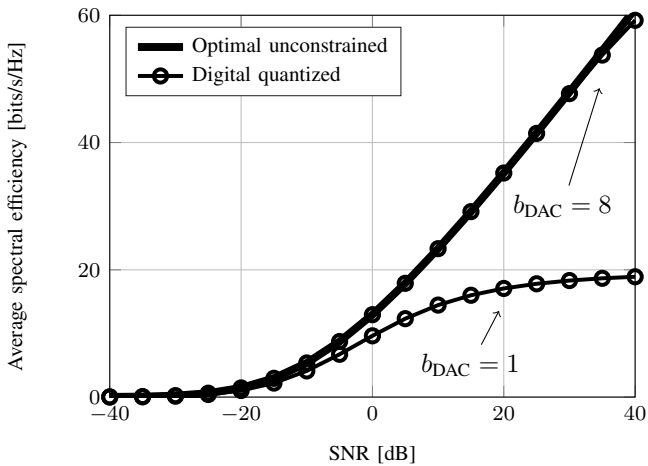
\begin{figure}
		\centering 
%
%
\begin{tikzpicture}

\begin{axis}[%
width=2.75in,
height=2in,
scale only axis,
xmin=-40,
xmax=40,
xlabel={SNR [dB]},
ymin=0,
ymax=60,
label style={font=\footnotesize},
tick label style={font=\footnotesize},
ylabel={Average spectral efficiency [bits/s/Hz]},
axis background/.style={fill=white},
xmajorgrids,
ymajorgrids,
legend style={at={(0.03,0.97)}, font=\footnotesize, anchor=north west, legend cell align=left, align=left, draw=white!15!black}
]
\addplot [color=black, line width=3.0pt]
  table[row sep=crcr]{%
-40	0.023951778078052\\
-35	0.0743672280951965\\
-30	0.222730245290796\\
-25	0.610048302311604\\
-20	1.4392203705603\\
-15	2.94958421627107\\
-10	5.36366152568404\\
-5	8.73032000612227\\
0	12.9493047746789\\
5	17.880764386662\\
10	23.3252573222168\\
15	29.1499313407969\\
20	35.2371105055864\\
25	41.5143414450976\\
30	47.9271825761503\\
35	54.4337266443846\\
40	60.9940824618635\\
};
\addlegendentry{Optimal unconstrained}

\addplot [color=black, line width=1.5pt, mark size=2.5pt, mark=o, mark options={solid, black}]
  table[row sep=crcr]{%
-40	0.0152934085658151\\
-35	0.0477802358302554\\
-30	0.145648806257813\\
-25	0.41541227215072\\
-20	1.03423134335176\\
-15	2.2020991285544\\
-10	4.11410000995189\\
-5	6.72404392042\\
0	9.62832377153281\\
5	12.3236736731554\\
10	14.4582289168388\\
15	16.0019565251775\\
20	17.0721426177748\\
25	17.8040793830407\\
30	18.3122369439721\\
35	18.6673722713106\\
40	18.9057989802836\\
};
\addlegendentry{Digital quantized}

\node[anchor=west] (source1) at (axis cs:6,5){$b_\text{DAC}=1$};
\node (destination1) at (axis cs:20, 15){};
\draw[->](source1)--(destination1);

\node[anchor=west] (source2) at (axis cs:20, 30){$b_\text{DAC}=8$};
\node (destination2) at (axis cs:35, 50){};
\draw[->](source2)--(destination2);

\addplot [color=black, line width=1.5pt, mark size=2.5pt, mark=o, mark options={solid, black}, forget plot]
  table[row sep=crcr]{%
-40	0.0239507918972543\\
-35	0.0743642183623391\\
-30	0.222721665697362\\
-25	0.61002740061499\\
-20	1.43917847056197\\
-15	2.94950820709141\\
-10	5.36353441101803\\
-5	8.73010574878541\\
0	12.9488885168968\\
5	17.8797815670407\\
10	23.3225532576668\\
15	29.1418619407036\\
20	35.212214533969\\
25	41.437123277789\\
30	47.6912338802234\\
35	53.7478618026967\\
40	59.2108406752189\\
};
\end{axis}
\end{tikzpicture}%
		\caption{Spectral efficiency of digital precoders. $N_t = 64$, $N_r = 4$, $L_t = N_s = 4$, $b_\text{DAC} \in \{1,8\}$.}
		\label{fig:f1}
	\end{figure}

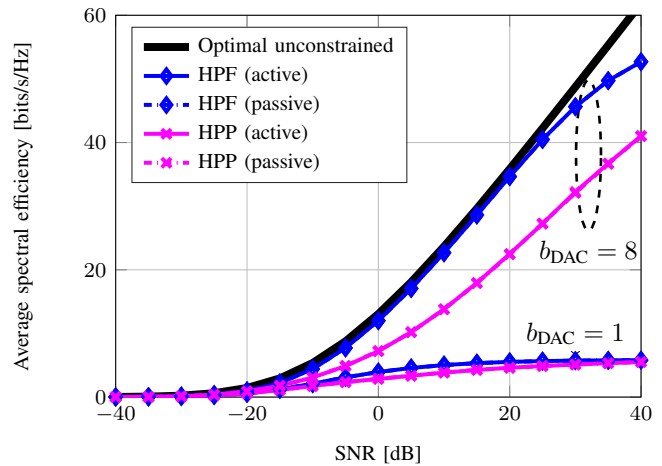
\begin{figure}
		\centering 
%
%
\definecolor{mycolor1}{rgb}{1.00000,0.00000,1.00000}%
\begin{tikzpicture}

\begin{axis}[%
width=2.75in,
height=2in,
scale only axis,
xmin=-40,
xmax=40,
xlabel={SNR [dB]},
ymin=0,
ymax=60,
ylabel={Average spectral efficiency [bits/s/Hz]},
label style={font=\footnotesize},
tick label style={font=\footnotesize},
axis background/.style={fill=white},
xmajorgrids,
ymajorgrids,
legend style={font=\footnotesize, at={(0.03,0.97)}, anchor=north west, legend cell align=left, align=left, draw=white!15!black}
]

\node (source2) at (axis cs:30, 10){$b_\text{DAC}=1$};
\node (destination2) at (axis cs:30, 5){};
\draw[->](source2)--(destination2);

\node (p8) at (axis cs:32 , 38) [ellipse, dashed, draw, label={below:$b_\text{DAC}=8$}, minimum width= 1mm, minimum height=20mm, line width = 1pt]  {};

\addplot [color=black, line width=3.0pt]
  table[row sep=crcr]{%
-40	0.0239254823085248\\
-35	0.0742925802918689\\
-30	0.222559627754027\\
-25	0.609378950151187\\
-20	1.43340805707738\\
-15	2.93478731797306\\
-10	5.35890369354337\\
-5	8.76227031140789\\
0	13.0578918424281\\
5	18.0757869880145\\
10	23.6638827044956\\
15	29.6230189224416\\
20	35.8506300029018\\
25	42.2492570554328\\
30	48.7483202167272\\
35	55.3033067778466\\
40	61.8976205965477\\
};
\addlegendentry{Optimal unconstrained}

\addplot [color=blue, line width=1.5pt, mark size=2.5pt, mark=diamond, mark options={solid, blue}]
  table[row sep=crcr]{%
-40	0.0131033828039731\\
-35	0.0376417668557641\\
-30	0.0983170379146131\\
-25	0.246946194957429\\
-20	0.581364095307626\\
-15	1.21357394390026\\
-10	2.11314215409945\\
-5	3.09419728646388\\
0	3.93613407712927\\
5	4.57724957325511\\
10	5.02927674617555\\
15	5.33208285738839\\
20	5.53001594590666\\
25	5.65526456870842\\
30	5.73108485207588\\
35	5.77696991218504\\
40	5.80542285749627\\
};
\addlegendentry{HPF (active)}

\addplot [color=blue, dashdotted, line width=1.5pt, mark size=2.5pt, mark=diamond, mark options={solid, blue}]
  table[row sep=crcr]{%
-40	0.0131033828039731\\
-35	0.0376417668557641\\
-30	0.0983170379146131\\
-25	0.246946194957429\\
-20	0.581364095307626\\
-15	1.21357394390026\\
-10	2.11314215409945\\
-5	3.09419728646388\\
0	3.93613407712927\\
5	4.57724957325511\\
10	5.02927674617555\\
15	5.33208285738839\\
20	5.53001594590666\\
25	5.65526456870842\\
30	5.73108485207588\\
35	5.77696991218504\\
40	5.80542285749627\\
};
\addlegendentry{HPF (passive)}

\addplot [color=mycolor1, line width=1.5pt, mark size=2.5pt, mark=x, mark options={solid, mycolor1}]
  table[row sep=crcr]{%
-40	0.00886074444596242\\
-35	0.0279246912556778\\
-30	0.0844779675827083\\
-25	0.239895180084811\\
-20	0.580680789572278\\
-15	1.11618754086009\\
-10	1.75112032706721\\
-5	2.34734910171214\\
0	2.85706372854228\\
5	3.34933609842472\\
10	3.84594588058902\\
15	4.26021910795364\\
20	4.61932276575106\\
25	4.88575241936699\\
30	5.10760407440171\\
35	5.32890010080579\\
40	5.50622883533624\\
};
\addlegendentry{HPP (active)}

\addplot [color=mycolor1, dashdotted, line width=1.5pt, mark size=2.5pt, mark=x, mark options={solid, mycolor1}]
  table[row sep=crcr]{%
-40	0.00884534636343095\\
-35	0.0276258873020932\\
-30	0.0850405956497775\\
-25	0.240725922354614\\
-20	0.579414461538246\\
-15	1.11902858097157\\
-10	1.75074324333124\\
-5	2.33986654807337\\
0	2.8732526705242\\
5	3.35366045277711\\
10	3.85653633551176\\
15	4.28981926201513\\
20	4.58058618380715\\
25	4.88209289407309\\
30	5.11327110370743\\
35	5.31667144326087\\
40	5.51610274725886\\
};
\addlegendentry{HPP (passive)}

\addplot [color=blue, line width=1.5pt, mark size=2.5pt, mark=diamond, mark options={solid, blue}, forget plot]
  table[row sep=crcr]{%
-40	0.0206344286669754\\
-35	0.059521958664985\\
-30	0.156734588147819\\
-25	0.400705051190793\\
-20	0.981270080663565\\
-15	2.21864871851617\\
-10	4.40832844349901\\
-5	7.72576492820258\\
0	12.008217136575\\
5	17.0667753714892\\
10	22.6753965647921\\
15	28.6157078409535\\
20	34.6525012909123\\
25	40.4792918177679\\
30	45.6325940361279\\
35	49.7292525633891\\
40	52.700534484071\\
};
\addplot [color=blue, dashdotted, line width=1.5pt, mark size=2.5pt, mark=diamond, mark options={solid, blue}, forget plot]
  table[row sep=crcr]{%
-40	0.0206344286669754\\
-35	0.059521958664985\\
-30	0.156734588147819\\
-25	0.400705051190793\\
-20	0.981270080663565\\
-15	2.21864871851617\\
-10	4.40832844349901\\
-5	7.72576492820258\\
0	12.008217136575\\
5	17.0667753714892\\
10	22.6753965647921\\
15	28.6157078409535\\
20	34.6525012909123\\
25	40.4792918177679\\
30	45.6325940361279\\
35	49.7292525633891\\
40	52.700534484071\\
};
\addplot [color=mycolor1, line width=1.5pt, mark size=2.5pt, mark=x, mark options={solid, mycolor1}, forget plot]
  table[row sep=crcr]{%
-40	0.0139340917232325\\
-35	0.0432224849310807\\
-30	0.132272780405471\\
-25	0.37045324247407\\
-20	0.892483936456921\\
-15	1.77497748311803\\
-10	3.07183000058579\\
-5	4.87926968337021\\
0	7.20750239355805\\
5	10.2004168486321\\
10	13.8085829912626\\
15	17.952517546854\\
20	22.4953132197671\\
25	27.2505224124725\\
30	32.2369268483114\\
35	36.7130504673835\\
40	41.0398972338826\\
};
\addplot [color=mycolor1, dashdotted, line width=1.5pt, mark size=2.5pt, mark=x, mark options={solid, mycolor1}, forget plot]
  table[row sep=crcr]{%
-40	0.0139407538526967\\
-35	0.0436975489409754\\
-30	0.132449272803262\\
-25	0.369138419011485\\
-20	0.888838442939613\\
-15	1.77729310361579\\
-10	3.07641779100637\\
-5	4.90808517656715\\
0	7.24741800020793\\
5	10.2010027835964\\
10	13.8020295705153\\
15	17.9052360623681\\
20	22.4413403351215\\
25	27.273530790852\\
30	32.0803908611855\\
35	36.6328962923158\\
40	40.9959524422028\\
};
\end{axis}
\end{tikzpicture}%
		\caption{Spectral efficiency of hybrid precoders (ignoring RF hardware losses). $N_t = 64$, $N_r = 4$, $L_t = N_s = 4$, $b_\text{DAC} \in \{1,8\}$.}
		\label{fig:f2}
	\end{figure}

\begin{figure}
		\centering 
%
%
\definecolor{mycolor1}{rgb}{1.00000,0.00000,1.00000}%
\begin{tikzpicture}

\begin{axis}[%
width=2.75in,
height=2in,
scale only axis,
xmin=-40,
xmax=40,
xlabel={SNR [dB]},
ymin=0,
ymax=60,
ylabel={Average spectral efficiency [bits/s/Hz]},
label style={font=\footnotesize},
tick label style={font=\footnotesize},
axis background/.style={fill=white},
xmajorgrids,
ymajorgrids,
legend style={font=\footnotesize, at={(0.03,0.97)}, anchor=north west, legend cell align=left, align=left, draw=white!15!black}
]

\node (p8) at (axis cs:32 , 29) [ellipse, dashed, draw, label={below:$b_\text{DAC}=8$}, minimum width= 2.5mm, minimum height=20mm, line width = 1pt]  {};

\node (source2) at (axis cs:30, 2){$b_\text{DAC}=1$};
\node (destination2) at (axis cs:30, 5){};
\draw[->](source2)--(destination2);

\addplot [color=black, line width=3.0pt]
  table[row sep=crcr]{%
-40	0.023951778078052\\
-35	0.0743672280951965\\
-30	0.222730245290796\\
-25	0.610048302311604\\
-20	1.4392203705603\\
-15	2.94958421627107\\
-10	5.36366152568404\\
-5	8.73032000612227\\
0	12.9493047746789\\
5	17.880764386662\\
10	23.3252573222168\\
15	29.1499313407969\\
20	35.2371105055864\\
25	41.5143414450976\\
30	47.9271825761503\\
35	54.4337266443846\\
40	60.9940824618635\\
};
\addlegendentry{Optimal unconstrained}

\addplot [color=blue, line width=1.5pt, mark size=2.5pt, mark=diamond, mark options={solid, blue}]
  table[row sep=crcr]{%
-40	0.0018554399122492\\
-35	0.0054272733138487\\
-30	0.0147090377236258\\
-25	0.0392785227219817\\
-20	0.104605074544108\\
-15	0.272057182936864\\
-10	0.662238144696444\\
-5	1.37467226024463\\
0	2.33581777685135\\
5	3.28815211800497\\
10	4.0598163198941\\
15	4.62319699241951\\
20	5.01612585416571\\
25	5.28777350247592\\
30	5.47519205246135\\
35	5.60515948689535\\
40	5.69375451574924\\
};
\addlegendentry{HPF (active)}

\addplot [color=blue, dashdotted, line width=1.5pt, mark size=2.5pt, mark=diamond, mark options={solid, blue}]
  table[row sep=crcr]{%
-40	0.00014420994805593\\
-35	0.000422718667047107\\
-30	0.00115079227554007\\
-25	0.00310550498566055\\
-20	0.00849672157943945\\
-15	0.0236475732842955\\
-10	0.0674609579769394\\
-5	0.188616577750944\\
0	0.497444348874146\\
5	1.12998421650111\\
10	2.06894210217931\\
15	3.06252035560172\\
20	3.89427401805687\\
25	4.50913839803859\\
30	4.93862217963644\\
35	5.23475649502951\\
40	5.4390322292997\\
};
\addlegendentry{HPF (passive)}

\addplot [color=mycolor1, line width=1.5pt, mark size=2.5pt, mark=x, mark options={solid, mycolor1}]
  table[row sep=crcr]{%
-40	0.00825374259968656\\
-35	0.0258951531534506\\
-30	0.0793605735745042\\
-25	0.227738243618425\\
-20	0.555640188109987\\
-15	1.07089242393123\\
-10	1.71558105164033\\
-5	2.35671912930043\\
0	2.92467417440188\\
5	3.43707725649282\\
10	3.84968924055201\\
15	4.18879126750067\\
20	4.54988400320245\\
25	4.82756428423713\\
30	5.07472999646393\\
35	5.23784230395404\\
40	5.41365477693278\\
};
\addlegendentry{HPP (active)}

\addplot [color=mycolor1, dashdotted, line width=1.5pt, mark size=2.5pt, mark=x, mark options={solid, mycolor1}]
  table[row sep=crcr]{%
-40	0.000645992542909897\\
-35	0.00204512793782061\\
-30	0.00638653462059747\\
-25	0.0199616864943469\\
-20	0.0577670930805537\\
-15	0.149775716872248\\
-10	0.350894844843921\\
-5	0.726385505227515\\
0	1.28816383972643\\
5	1.90975095426187\\
10	2.48633431877334\\
15	2.99388026601888\\
20	3.48281046578092\\
25	3.90798508093362\\
30	4.28829025334977\\
35	4.62040881615968\\
40	4.88711538317409\\
};
\addlegendentry{HPP (passive)}

\addplot [color=blue, line width=1.5pt, mark size=2.5pt, mark=diamond, mark options={solid, blue}, forget plot]
  table[row sep=crcr]{%
-40	0.00291550544491119\\
-35	0.00853306731829346\\
-30	0.0231554053829977\\
-25	0.0620181408348968\\
-20	0.166447508130959\\
-15	0.441516703664304\\
-10	1.12686819776112\\
-5	2.56915022582319\\
0	5.07866686877564\\
5	8.67087856202795\\
10	13.1678080482948\\
15	18.3236167510721\\
20	23.9383621156543\\
25	29.8170548961375\\
30	35.7295698457241\\
35	41.3266671082241\\
40	46.1482114768541\\
};
\addplot [color=blue, dashdotted, line width=1.5pt, mark size=2.5pt, mark=diamond, mark options={solid, blue}, forget plot]
  table[row sep=crcr]{%
-40	0.000226528231619288\\
-35	0.000664046548974264\\
-30	0.00180795139493924\\
-25	0.00488005849227511\\
-20	0.0133604246879784\\
-15	0.0372492667717424\\
-10	0.106801764336563\\
-5	0.302761788697736\\
0	0.827814742847785\\
5	2.03639473371787\\
10	4.28831834288359\\
15	7.66787904155717\\
20	12.0157493647663\\
25	17.0815605703365\\
30	22.6360103408013\\
35	28.4882257778875\\
40	34.4270515596615\\
};
\addplot [color=mycolor1, line width=1.5pt, mark size=2.5pt, mark=x, mark options={solid, mycolor1}, forget plot]
  table[row sep=crcr]{%
-40	0.0129346298787272\\
-35	0.0406800554638608\\
-30	0.124832935208443\\
-25	0.353217264469351\\
-20	0.846346425262916\\
-15	1.68974151697218\\
-10	2.99821278452603\\
-5	4.82659609191217\\
0	7.29414939734206\\
5	10.2696802569089\\
10	13.8573118662899\\
15	17.8594538379809\\
20	22.2989151460178\\
25	27.1069570441851\\
30	31.8315811542816\\
35	36.3132776568641\\
40	40.2652935524307\\
};
\addplot [color=mycolor1, dashdotted, line width=1.5pt, mark size=2.5pt, mark=x, mark options={solid, mycolor1}, forget plot]
  table[row sep=crcr]{%
-40	0.00102118916679971\\
-35	0.00321915041273817\\
-30	0.0101571620706556\\
-25	0.0312906563139801\\
-20	0.0898364582989358\\
-15	0.232477012189754\\
-10	0.550852245062465\\
-5	1.18160211655802\\
0	2.29390153569869\\
5	3.96365061001391\\
10	6.24595076463206\\
15	9.0883940264314\\
20	12.5635200061062\\
25	16.5134818071773\\
30	20.9800180089358\\
35	25.7566998050107\\
40	30.4989348147065\\
};
\end{axis}
\end{tikzpicture}%
		\caption{Spectral efficiency of hybrid precoders (considering RF hardware losses). $N_t = 64$, $N_r = 4$, $L_t = N_s = 4$, $b_\text{DAC} \in \{1,8\}$.}
		\label{fig:f3}
	\end{figure}
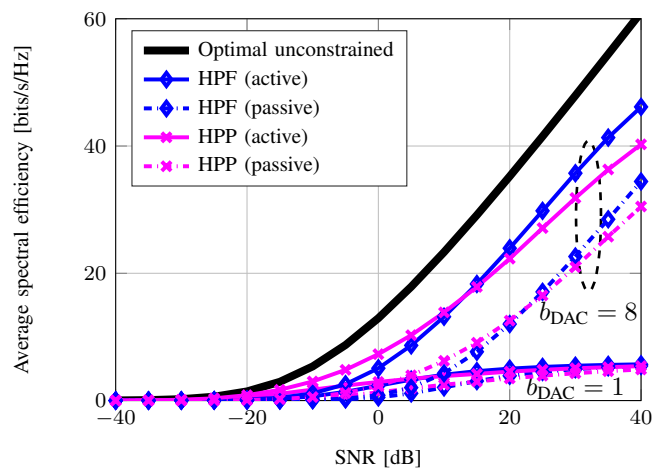

	Figures \ref{fig:f1}, \ref{fig:f2}, and \ref{fig:f3} show the average spectral efficiency of the digital and hybrid precoders as a function of the SNR for DAC resolution of $1$ and $8$ bits. The curves depicted in Figure~\ref{fig:f1} indicate that the quantized digital precoder does not lose much spectral efficiency at low SNR, whereas its performance saturates at high SNR. This is because quantization distortion does not drop at high SNR, as discussed in Section~\ref{sec:achrate}. Since the distortion caused by $1$-bit DACs is important, the spectral efficiency already saturates at $20\,\si{\decibel}$, whereas it occurs only at $>40\,\si{\decibel}$ for $8$-bit DACs. The spectral efficiency of lossless hybrid precoders using active and passive phase-shifters is compared to that of the unquantized digital precoder in Figure~\ref{fig:f2}. As observed in several hybrid precoding works, HPF outperforms HPP when RF hardware losses are ignored. In this case, the type of phase-shifting circuit does not have any impact on the spectral efficiency as no insertion losses are considered yet. Comparing Figures~\ref{fig:f1} and \ref{fig:f2}, one recognizes that hybrid precoding saturates at lower data rates compared to digital precoding. This is because the quantization distortion $\mb{e}$ is filtered by analog precoding matrix $\Frf$, as visible in Equation~\eqref{eq:precod_lowres}, increasing the total noise $\mb{n}'$ power. In general, digital precoding provides higher data rate than hybrid precoding for fixed number of bits per DAC and SNR.
	
	In Figure~\ref{fig:f3}, the average spectral efficiency of the hybrid precoders is shown considering RF hardware losses. In this scenario, HPF performs similarly to HPP because its insertion loss is larger than that of the other architecture, reducing the effective transmit power and, consequently, the spectral efficiency. When RF losses are taken into account, the phase-shifting implementation becomes important. As  shown in Figure~\ref{fig:f3}, precoders with active phase-shifters are more spectral-efficient since these components do not attenuate the signals like passive phase-shifters, however they come with increased power consumption. We observe in Figure~\ref{fig:f3} SNR losses of $10\,\si{\decibel}$ for the active PSN and $20\,\si{\decibel}$ for the passive PSN when $b_\text{DAC}=8$. Similar losses have been reported in other works \cite{garcia-rodriguez_hybrid_2016,zochmann_comparing_2016} for hybrid precoders with fully-connected PSN. Such important power losses occur due to the large number of lossy RF components employed in this mmWave massive MIMO setup. 
	
	~
	~
	
	\subsection{Power Consumption} 
	
\begin{figure}
		\centering 
%
%
\definecolor{mycolor1}{rgb}{1.00000,0.00000,1.00000}%
\begin{tikzpicture}

\begin{axis}[%
width=1.25in,
height=2in,
at={(0in,0in)},
title={Static},
title style={font=\footnotesize},
scale only axis,
xmin=0,
xmax=600,
xlabel={$N_t$},
ymode=log,
ymin=0.001,
ymax=10000,
yminorticks=true,
label style={font=\footnotesize},
tick label style={font=\footnotesize},
ylabel={Power consumption [W]},
axis background/.style={fill=white},
xmajorgrids,
ymajorgrids,
yminorgrids,
legend style={at={(0.6,0.025)}, anchor=south,font=\tiny, legend cell align=left, align=left, draw=white!15!black}
]
\addplot [color=black, line width=1.5pt, mark size=2.5pt, mark=o, mark options={solid, black}]
  table[row sep=crcr]{%
32	5.3153237037037\\
64	6.9044437037037\\
128	10.0826837037037\\
256	16.4391637037037\\
512	29.1521237037037\\
};
\addlegendentry{Digital quant.}

\addplot [color=blue, line width=1.5pt, mark size=2.5pt, mark=diamond, mark options={solid, blue}]
  table[row sep=crcr]{%
32	2.8802764040636\\
64	6.27390205356398\\
128	22.7370806653223\\
256	89.3662799238767\\
512	355.534769893208\\
};
\addlegendentry{HPF (active)}

\addplot [color=blue, dashdotted, line width=1.5pt, mark size=2.5pt, mark=diamond, mark options={solid, blue}]
  table[row sep=crcr]{%
32	0.228635981597282\\
64	0.261750303560859\\
128	0.435219606795198\\
256	0.822929974786893\\
512	1.61385170217065\\
};
\addlegendentry{HPF (passive)}

\addplot [color=mycolor1, line width=1.5pt, mark size=2.5pt, mark=x, mark options={solid, mycolor1}]
  table[row sep=crcr]{%
32	4.43241711465115\\
64	5.22293711465115\\
128	6.80397711465115\\
256	9.96605711465115\\
512	16.2902171146511\\
};
\addlegendentry{HPP (active)}

\addplot [color=mycolor1, dashdotted, line width=1.5pt, mark size=2.5pt, mark=x, mark options={solid, mycolor1}]
  table[row sep=crcr]{%
32	0.402774657418216\\
64	0.502094657418216\\
128	0.700734657418216\\
256	1.09801465741822\\
512	1.89257465741822\\
};
\addlegendentry{HPP (passive)}

\addplot [color=black, line width=1.5pt, mark size=2.5pt, mark=o, mark options={solid, black}, forget plot]
  table[row sep=crcr]{%
32	6.4730837037037\\
64	9.2199637037037\\
128	14.7137237037037\\
256	25.7012437037037\\
512	47.6762837037037\\
};
\addplot [color=blue, line width=1.5pt, mark size=2.5pt, mark=diamond, mark options={solid, blue}, forget plot]
  table[row sep=crcr]{%
32	2.9526364040636\\
64	6.41862205356398\\
128	23.0265206653223\\
256	89.9451599238767\\
512	356.692529893208\\
};
\addplot [color=blue, dashdotted, line width=1.5pt, mark size=2.5pt, mark=diamond, mark options={solid, blue}, forget plot]
  table[row sep=crcr]{%
32	0.300995981597282\\
64	0.406470303560859\\
128	0.724659606795198\\
256	1.40180997478689\\
512	2.77161170217065\\
};
\addplot [color=mycolor1, line width=1.5pt, mark size=2.5pt, mark=x, mark options={solid, mycolor1}, forget plot]
  table[row sep=crcr]{%
32	4.50477711465115\\
64	5.36765711465115\\
128	7.09341711465115\\
256	10.5449371146512\\
512	17.4479771146512\\
};
\addplot [color=mycolor1, dashdotted, line width=1.5pt, mark size=2.5pt, mark=x, mark options={solid, mycolor1}, forget plot]
  table[row sep=crcr]{%
32	0.475134657418216\\
64	0.646814657418216\\
128	0.990174657418216\\
256	1.67689465741822\\
512	3.05033465741822\\
};
\addplot [color=black, line width=1.5pt, mark size=2.5pt, mark=o, mark options={solid, black}, forget plot]
  table[row sep=crcr]{%
32	7.6481237037037\\
64	11.5700437037037\\
128	19.4138837037037\\
256	35.1015637037037\\
512	66.4769237037037\\
};
\addplot [color=blue, line width=1.5pt, mark size=2.5pt, mark=diamond, mark options={solid, blue}, forget plot]
  table[row sep=crcr]{%
32	3.0260764040636\\
64	6.56550205356398\\
128	23.3202806653223\\
256	90.5326799238767\\
512	357.867569893208\\
};
\addplot [color=blue, dashdotted, line width=1.5pt, mark size=2.5pt, mark=diamond, mark options={solid, blue}, forget plot]
  table[row sep=crcr]{%
32	0.374435981597282\\
64	0.553350303560859\\
128	1.0184196067952\\
256	1.98932997478689\\
512	3.94665170217065\\
};
\addplot [color=mycolor1, line width=1.5pt, mark size=2.5pt, mark=x, mark options={solid, mycolor1}, forget plot]
  table[row sep=crcr]{%
32	4.57821711465115\\
64	5.51453711465115\\
128	7.38717711465115\\
256	11.1324571146512\\
512	18.6230171146512\\
};
\addplot [color=mycolor1, dashdotted, line width=1.5pt, mark size=2.5pt, mark=x, mark options={solid, mycolor1}, forget plot]
  table[row sep=crcr]{%
32	0.548574657418216\\
64	0.793694657418216\\
128	1.28393465741822\\
256	2.26441465741822\\
512	4.22537465741822\\
};
\addplot [color=black, line width=1.5pt, mark size=2.5pt, mark=o, mark options={solid, black}, forget plot]
  table[row sep=crcr]{%
32	8.8922837037037\\
64	14.0583637037037\\
128	24.3905237037037\\
256	45.0548437037037\\
512	86.3834837037037\\
};
\addplot [color=blue, line width=1.5pt, mark size=2.5pt, mark=diamond, mark options={solid, blue}, forget plot]
  table[row sep=crcr]{%
32	3.1038364040636\\
64	6.72102205356398\\
128	23.6313206653223\\
256	91.1547599238767\\
512	359.111729893208\\
};
\addplot [color=blue, dashdotted, line width=1.5pt, mark size=2.5pt, mark=diamond, mark options={solid, blue}, forget plot]
  table[row sep=crcr]{%
32	0.452195981597282\\
64	0.708870303560859\\
128	1.3294596067952\\
256	2.61140997478689\\
512	5.19081170217065\\
};
\addplot [color=mycolor1, line width=1.5pt, mark size=2.5pt, mark=x, mark options={solid, mycolor1}, forget plot]
  table[row sep=crcr]{%
32	4.65597711465115\\
64	5.67005711465115\\
128	7.69821711465115\\
256	11.7545371146512\\
512	19.8671771146512\\
};
\addplot [color=mycolor1, dashdotted, line width=1.5pt, mark size=2.5pt, mark=x, mark options={solid, mycolor1}, forget plot]
  table[row sep=crcr]{%
32	0.626334657418216\\
64	0.949214657418216\\
128	1.59497465741822\\
256	2.88649465741822\\
512	5.46953465741822\\
};
\end{axis}

\begin{axis}[%
width=1.25in,
height=2in,
at={(1.45in,0in)},
scale only axis,
title={Computational},
title style={font=\footnotesize},
xmin=0,
xmax=600,
xlabel={$N_t$},
ymode=log,
ymin=0.001,
ymax=10000,
ymajorticks=false,
yminorticks=true,
axis background/.style={fill=white},
label style={font=\footnotesize},
tick label style={font=\footnotesize},
xmajorgrids,
ymajorgrids,
yminorgrids
]
\addplot [color=black, line width=1.5pt, mark size=2.5pt, mark=o, mark options={solid, black}, forget plot]
  table[row sep=crcr]{%
32	0.62754928\\
64	5.00972768\\
128	40.05648832\\
256	320.40924032\\
512	2563.18859008\\
};
\addplot [color=blue, line width=1.5pt, mark size=2.5pt, mark=diamond, mark options={solid, blue}, forget plot]
  table[row sep=crcr]{%
32	0.9134630875\\
64	7.26170516\\
128	57.93230956\\
256	462.85848248\\
512	3700.5585496\\
};
\addplot [color=mycolor1, line width=1.5pt, mark size=2.5pt, mark=x, mark options={solid, mycolor1}, forget plot]
  table[row sep=crcr]{%
32	0.020578787265625\\
64	0.042648011015625\\
128	0.095869701015625\\
256	0.264312461015625\\
512	1.05452678101562\\
};
\end{axis}
\end{tikzpicture}%
		\caption{$N_r=4$, $L_t=N_s$, $b_\text{DAC} \in \{1,3,5,7\}$, $(N_t, N_s) \in \{(32,2), (64, 4), (128, 8), (256, 16), (512,32)\}$.}
		\label{fig:f4}
	\end{figure}
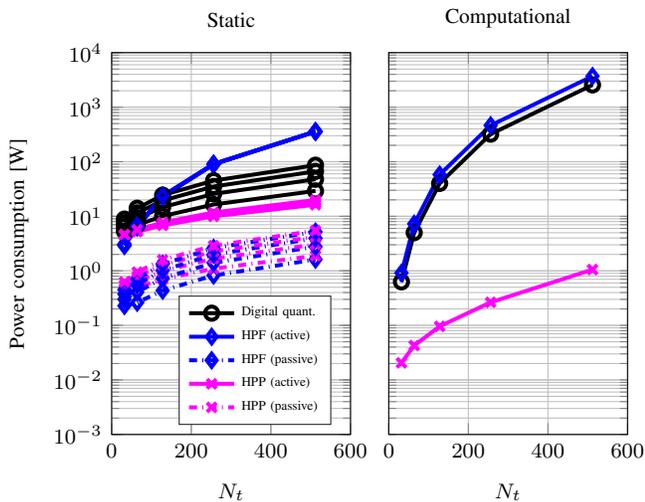

	The static power consumption is plotted as a function of the number $N_t$ of antennas for $b_\text{DAC} \in \{1, 3, 5, 7\}$ in the left part of Figure~\ref{fig:f4}. Despite the reduced number of RF chains, active HPF can still be power hungry, even more than the digital precoder. This is mainly due to the power consumption of the fully-connected PSN. To solve this issue, one can employ passive phase-shifters, drastically decreasing the power demand of HPF, as shown in Figure~\ref{fig:f4}. However, as observed in Figure~\ref{fig:f3}, the use of such passive components incurs in spectral efficiency degradation. Interestingly, passive HPP consumes more power than passive HPF. This is because the PAs in the former scheme consume more power than in the latter, as partially-connected PSNs lose less power, hence delivering more power to the amplifiers. One can also observe in Figure~\ref{fig:f4} the effect of DAC resolution on the static power consumption. Digital precoding is more sensitive to DAC resolution because it uses $N_t$ DACs, while hybrid precoding uses only $L_t$.
	
	In practice, computational power consumption plays an important role on system energy efficiency. The plot in the right part of Figure~\ref{fig:f4} suggests that it is negligible when the antenna array is comparatively small (up to $64$ antennas). By contrast, it becomes substantial for HPF and digital precoder with growing $N_t$. HPP is not as power hungry since the number of flops necessary to run Algorithm~\ref{alg:ppsn} scales linearly with the number $N_t$ of transmit antennas, whereas the other schemes contain quadratic and cubic terms of $N_t$. We stress that the proposed power consumption model mainly serves to provide energy efficiency estimates, allowing us to compare different precoding structures. In practice, optimized hardware and software designs may reduce the values provided by our model. For example, one can simplify PA, DAC, and modulation design when employing $1$-bit quantization. However, considering such specialized implementations is out of the scope of this work.
	
	\subsection{Energy Efficiency} 
	
\begin{figure}
		\centering 
%
%
\definecolor{mycolor1}{rgb}{1.00000,0.00000,1.00000}%
\begin{tikzpicture}

\begin{axis}[%
width=2.5in,
height=2in,
scale only axis,
xmode=log,
xmin=0.0,
xmax=100,
xminorticks=true,
xlabel={Spectral efficiency [bits/s/Hertz]},
ymode=log,
ymin=0.0,
ymax=10,
yminorticks=true,
ylabel={Energy efficiency [bits/J]},
label style={font=\footnotesize},
tick label style={font=\footnotesize},
axis background/.style={fill=white},
xmajorgrids,
xminorgrids,
ymajorgrids,
yminorgrids,
legend style={font=\footnotesize, at={(0.97,0.62)}, anchor=south east, legend cell align=left, align=left, draw=white!15!black, font=\tiny}
]
\addplot [color=black, line width=1.5pt, mark size=2pt, mark=o, mark options={solid, black}]
  table[row sep=crcr]{%
9.68756418667611	1.403091197844\\
11.884861553417	1.47449667906303\\
12.6481043832306	1.37181715565212\\
12.9013980050671	1.24203292138349\\
12.9814882794947	1.12199129164388\\
13.0097709444447	1.01770153160014\\
13.0147664745711	0.925766806782949\\
13.0160158002188	0.842126787397531\\
};
\addlegendentry{Digital quantized}

\addplot [color=blue, line width=1.5pt, mark size=2pt, mark=diamond, mark options={solid, blue}]
  table[row sep=crcr]{%
2.35299630200324	0.375045112581346\\
3.92434075982189	0.61838211730826\\
4.68378518848432	0.729718177733119\\
4.97487605044471	0.766358032509721\\
5.07214185775613	0.772544401993263\\
5.10715605477337	0.768994581755157\\
5.1133780290021	0.760803637936377\\
5.11493585260453	0.751270391755853\\
};
\addlegendentry{HPF (active)}

\addplot [color=blue, dashdotted, line width=1.5pt, mark size=2pt, mark=+, mark options={solid, blue}]
  table[row sep=crcr]{%
0.500664969863147	1.91275793400079\\
0.71954740735653	2.15439609978203\\
0.796492288258594	1.95953377474559\\
0.821466170817365	1.71342146025421\\
0.82927716598574	1.49864771131283\\
0.83202459115465	1.32237351155265\\
0.832509255013869	1.17441688674492\\
0.832630435063571	1.04571558171037\\
};
\addlegendentry{HPF (passive)}

\addplot [color=mycolor1, line width=1.5pt, mark size=2pt, mark=x, mark options={solid, mycolor1}]
  table[row sep=crcr]{%
2.94370029842472	0.563610136175521\\
5.04153835933078	0.952100043147077\\
6.35927526400514	1.18473947351207\\
7.01372500831627	1.28914144489765\\
7.27446396520854	1.3191431690398\\
7.39175223022011	1.32222783519343\\
7.43370424665042	1.31104574369138\\
7.42989843841397	1.29049160247688\\
};
\addlegendentry{HPP (active)}

\addplot [color=mycolor1, dashdotted, line width=1.5pt, mark size=2pt, mark=square, mark options={solid, mycolor1}]
  table[row sep=crcr]{%
1.31319230955153	2.61542776874783\\
1.95626575592211	3.40614262199679\\
2.19074603579562	3.38697648649469\\
2.29471301772936	3.1880992114398\\
2.32236567882207	2.92601903908036\\
2.32971000301359	2.67926066331946\\
2.33504130542983	2.45997181689223\\
2.3439946582136	2.26128879520531\\
};
\addlegendentry{HPP (passive)}

\end{axis}
\end{tikzpicture}%
		\caption{Energy-spectral efficiency curves for varying DAC resolution. $\text{SNR}=0\,\si{\decibel}$, $N_t= 64$, $N_r=4$, $L_t = N_s =4$, $b_\text{DAC} \in \{1,\ldots,8 \}$.}
		\label{fig:f5}
	\end{figure}
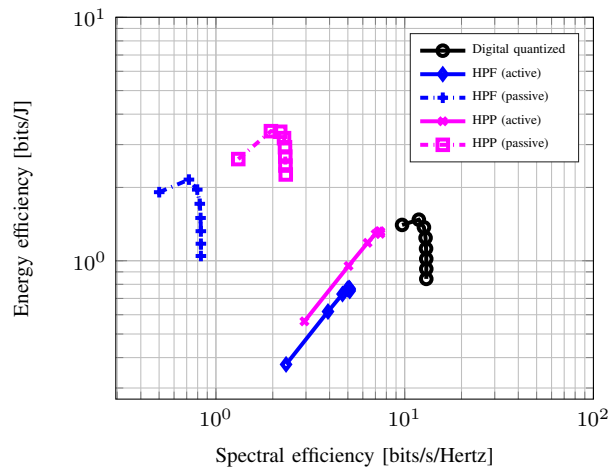

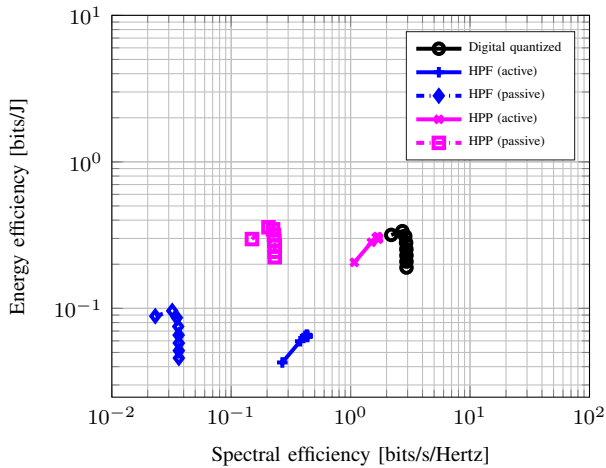
\begin{figure}
		\centering 
%
%
\definecolor{mycolor1}{rgb}{1.00000,0.00000,1.00000}%
\begin{tikzpicture}

\begin{axis}[%
width=2.5in,
height=2in,
scale only axis,
xmode=log,
xmin=0.0,
xmax=100,
xminorticks=true,
xlabel={Spectral efficiency [bits/s/Hertz]},
ymode=log,
ymin=0.0,
ymax=10,
yminorticks=true,
ylabel={Energy efficiency [bits/J]},
label style={font=\footnotesize},
tick label style={font=\footnotesize},
axis background/.style={fill=white},
xmajorgrids,
xminorgrids,
ymajorgrids,
yminorgrids,
legend style={font=\footnotesize, at={(0.97,0.62)}, anchor=south east, legend cell align=left, align=left, draw=white!15!black, font=\tiny}
]
\addplot [color=black, line width=1.5pt, mark size=2pt, mark=o, mark options={solid, black}]
  table[row sep=crcr]{%
2.19005822387771	0.317195463944895\\
2.71085484412592	0.336322509700283\\
2.8664064972077	0.310891299502215\\
2.91439423443933	0.280572197186863\\
2.92916707999305	0.253168194952918\\
2.9343370273897	0.229541265542471\\
2.93524763237281	0.208790133349553\\
2.93547524422636	0.189923120473146\\
};
\addlegendentry{Digital quantized}

\addplot [color=blue, line width=1.5pt, mark size=2pt, mark=+, mark options={solid, blue}]
  table[row sep=crcr]{%
0.268046359244609	0.0427240267629523\\
0.378709059820181	0.0596754778925093\\
0.416597814173009	0.0649045559461926\\
0.428778543170769	0.066051470909986\\
0.432576395162354	0.0658862630204398\\
0.433910895131532	0.0653348211298167\\
0.434146236743061	0.0645952703745176\\
0.434205075564941	0.0637750749221906\\
};
\addlegendentry{HPF (active)}

\addplot [color=blue, dashdotted, line width=1.5pt, mark size=2pt, mark=diamond, mark options={solid, blue}]
  table[row sep=crcr]{%
0.023155965690842	0.0884658599276774\\
0.0321093744892287	0.0961386427896036\\
0.0350944989129421	0.0863396381125482\\
0.0360455146074331	0.0751840556170798\\
0.0363411731165892	0.0656748046991762\\
0.0364449653791382	0.057923596681132\\
0.0364632641076816	0.0514385550142476\\
0.0364678388104188	0.0458006165393722\\
};
\addlegendentry{HPF (passive)}

\addplot [color=mycolor1, line width=1.5pt, mark size=2pt, mark=x, mark options={solid, mycolor1}]
  table[row sep=crcr]{%
1.07630465943698	0.206072682058104\\
1.49489629026285	0.282312802366259\\
1.63802787103409	0.305166264544554\\
1.67920250434885	0.308641918547601\\
1.69140193971868	0.306716938258503\\
1.70289638249245	0.304612076711164\\
1.70884357530527	0.301380310771421\\
1.70585934619912	0.296288997692758\\
};
\addlegendentry{HPP (active)}

\addplot [color=mycolor1, dashdotted, line width=1.5pt, mark size=2pt, mark=square, mark options={solid, mycolor1}]
  table[row sep=crcr]{%
0.149212436747917	0.297179893359493\\
0.205386882316637	0.357608372860353\\
0.224100505684737	0.346467883982782\\
0.228375873829257	0.317288017125285\\
0.230036338704041	0.289829768354897\\
0.231938669867082	0.266738844608844\\
0.231433300342488	0.243815556927836\\
0.231688874604227	0.22351392921499\\
};
\addlegendentry{HPP (passive)}

\end{axis}
\end{tikzpicture}%
		\caption{Energy-spectral efficiency curves for varying DAC resolution. $\text{SNR}=-15\,\si{\decibel}$, $N_t= 64$, $N_r=4$, $L_t = N_s =4$, $b_\text{DAC} \in \{1,\ldots,8 \}$.}
		\label{fig:f6}
	\end{figure}

	Numerical experiments were also carried out to study the energy efficiency of the proposed precoding strategies and its trade-off with spectral efficiency when the RF hardware losses are considered. The compromise between active and passive phase-shifters is investigated as well. Mathematically, we define the energy efficiency of a system as the ratio of its spectral efficiency to its static power consumption~\cite{abbas_millimeter_2016}. Figures~\ref{fig:f5} and~\ref{fig:f6} show the energy-spectral efficiency curves of different precoding schemes for varying DAC resolution at high and low SNR, respectively. A system designer should aim at maximizing both figures of merit, approaching the top right corner of the chart in Figs.~\ref{fig:f5} and~\ref{fig:f6}.
	
	In contrast to what is usually claimed by several hybrid precoding papers, Figures~\ref{fig:f5} and~\ref{fig:f6} indicate that HPF is not necessarily an efficient scheme. This is because the combination of large insertion losses due to phase-shifters and power combiners with the increased power consumption of fully-connected PSN degrades the energy efficiency. Such insertion losses are disregarded in most hybrid precoding works, leading to the erroneous conclusion that fully-connected PSN is the most spectral-efficient hybrid topology in general. As  an alternative to HPF, transmitters equipped with partially-connected PSNs lose less power by employing a reduced number of phase-shifters and no power combiners at all. Therefore, as observed in Figures~\ref{fig:f5} and~\ref{fig:f6}, HPP can be more energy- and spectral-efficient than HPF. These figures also reveal that phase-shifter topology offers an energy-spectral efficiency trade-off. As active phase-shifting amplifies the shifted signals, it favors spectral efficiency. On the other hand, passive phase-shifting offers negligible power consumption, promoting energy efficiency. Among the studied precoding methods, digital precoding is the most spectral-efficient solution for obvious reasons. Figure~\ref{fig:f6} indicates that digital precoding is as energy-efficient as HPP at low SNR. In this scenario, digital precoding compensates for the important power consumption by high data throughput, thus increasing energy efficiency. Regarding DAC resolution, Figures~\ref{fig:f5} and~\ref{fig:f6} also show that using $b_\text{DAC} > 3$ does not necessarily ameliorate energy efficiency. In fact, passive HPF and HPP and digital precoding become less efficient for higher bit resolution.
	
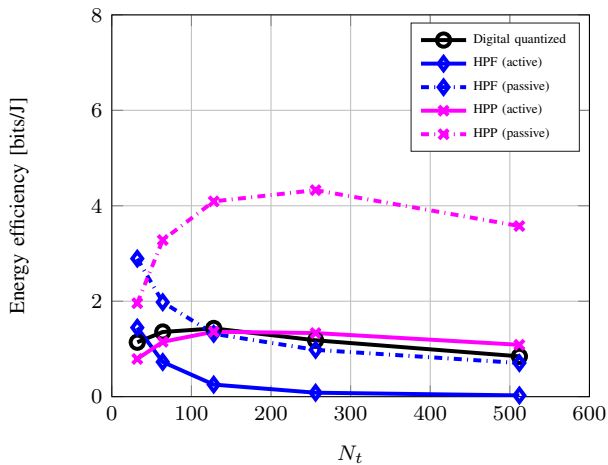
\begin{figure}
		\centering 
%
%
\definecolor{mycolor1}{rgb}{1.00000,0.00000,1.00000}%
\begin{tikzpicture}

\begin{axis}[%
width=2.5in,
height=2in,
scale only axis,
unbounded coords=jump,
xmin=0,
xmax=600,
xlabel={$N_t$},
ymin=0,
ymax=8,
ylabel={Energy efficiency [bits/J]},
axis background/.style={fill=white},
label style={font=\footnotesize},
tick label style={font=\footnotesize},
xmajorgrids,
ymajorgrids,
legend style={font=\tiny, legend cell align=left, align=left, draw=white!15!black}
]
\addplot [color=black, line width=1.5pt, mark size=2.5pt, mark=o, mark options={solid, black}]
  table[row sep=crcr]{%
32	1.14051319928343\\
64	1.35525951575322\\
128	1.42994031672392\\
256	1.18127376459713\\
512	0.844860809818244\\
};
\addlegendentry{Digital quantized}

\addplot [color=blue, line width=1.5pt, mark size=2.5pt, mark=diamond, mark options={solid, blue}]
  table[row sep=crcr]{%
32	1.44673460331712\\
64	0.727228290374747\\
128	0.253034796262405\\
256	0.0831641951838888\\
512	0.0264418466505556\\
};
\addlegendentry{HPF (active)}

\addplot [color=blue, dashdotted, line width=1.5pt, mark size=2.5pt, mark=diamond, mark options={solid, blue}]
  table[row sep=crcr]{%
32	2.89260037565181\\
64	1.9797957979937\\
128	1.31630416158842\\
256	0.981242434061543\\
512	0.705111066592511\\
};
\addlegendentry{HPF (passive)}

\addplot [color=mycolor1, line width=1.5pt, mark size=2.5pt, mark=x, mark options={solid, mycolor1}]
  table[row sep=crcr]{%
32	0.790280172237373\\
64	1.14887726701724\\
128	1.36067188802698\\
256	1.33254914656165\\
512	1.08666675876873\\
};
\addlegendentry{HPP (active)}

\addplot [color=mycolor1, dashdotted, line width=1.5pt, mark size=2.5pt, mark=x, mark options={solid, mycolor1}]
  table[row sep=crcr]{%
32	1.95866702389191\\
64	3.28131589988082\\
128	4.09196522730048\\
256	4.32733997149757\\
512	3.57511995602941\\
};
\addlegendentry{HPP (passive)}

\end{axis}
\end{tikzpicture}%
		\caption{$\text{SNR}=0\,\si{\decibel}$, $b_\text{DAC} =3$, $N_r = L_t=N_s$, and $(N_t, N_s) \in \{(32, 2), (64, 4), (128, 8), (256,16), (512,32)\}$.}
		\label{fig:f7}
	\end{figure}

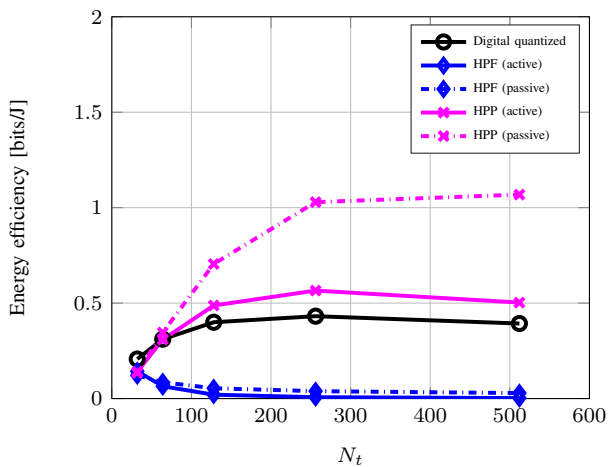
\begin{figure}
		\centering 
%
%
\definecolor{mycolor1}{rgb}{1.00000,0.00000,1.00000}%
\begin{tikzpicture}

\begin{axis}[%
width=2.5in,
height=2in,
scale only axis,
unbounded coords=jump,
xmin=0,
xmax=600,
xlabel={$N_t$},
ymin=0,
ymax=2,
ylabel={Energy efficiency [bits/J]},
label style={font=\footnotesize},
tick label style={font=\footnotesize},
axis background/.style={fill=white},
xmajorgrids,
ymajorgrids,
legend style={font=\tiny, legend cell align=left, align=left, draw=white!15!black}
]
\addplot [color=black, line width=1.5pt, mark size=2.5pt, mark=o, mark options={solid, black}]
  table[row sep=crcr]{%
32	0.207041001549858\\
64	0.311255780182271\\
128	0.399740065165548\\
256	0.431523964838507\\
512	0.393108694779208\\
};
\addlegendentry{Digital quantized}

\addplot [color=blue, line width=1.5pt, mark size=2.5pt, mark=diamond, mark options={solid, blue}]
  table[row sep=crcr]{%
32	0.143355884405989\\
64	0.0645373719009054\\
128	0.0203591852306439\\
256	0.0071381979581919\\
512	0.00257321122947059\\
};
\addlegendentry{HPF (active)}

\addplot [color=blue, dashdotted, line width=1.5pt, mark size=2.5pt, mark=diamond, mark options={solid, blue}]
  table[row sep=crcr]{%
32	0.121755894816773\\
64	0.0855416932017961\\
128	0.0539720028871514\\
256	0.0388244535747234\\
512	0.0287859049443862\\
};
\addlegendentry{HPF (passive)}

\addplot [color=mycolor1, line width=1.5pt, mark size=2.5pt, mark=x, mark options={solid, mycolor1}]
  table[row sep=crcr]{%
32	0.140543731654604\\
64	0.308770706999603\\
128	0.487539766223876\\
256	0.565656347729529\\
512	0.502881117632087\\
};
\addlegendentry{HPP (active)}

\addplot [color=mycolor1, dashdotted, line width=1.5pt, mark size=2.5pt, mark=x, mark options={solid, mycolor1}]
  table[row sep=crcr]{%
32	0.134887227114065\\
64	0.345967874559473\\
128	0.705097443778755\\
256	1.02885941033323\\
512	1.06841299683618\\
};
\addlegendentry{HPP (passive)}

\end{axis}
\end{tikzpicture}%
		\caption{$\text{SNR}=-15\,\si{\decibel}$, $b_\text{DAC} =3$, $N_r = L_t=N_s$, and $(N_t, N_s) \in \{(32, 2), (64, 4), (128, 8), (256,16), (512,32)\}$.}
		\label{fig:f8}
	\end{figure}

	Do the obtained energy efficiency results hold for other scenarios? For example, do they still hold for antenna arrays with more elements? To answer this question, consider Figures~\ref{fig:f7} and \ref{fig:f8}, where the energy efficiency is plotted for different $(N_t, N_s)$ tuples at high and low SNR, respectively. These plots were obtained by setting $b_\text{DAC} = 3$, motivated by the discussion above. In general, HPP is the most energy-efficient scheme, followed by digital precoding, and HPF at both high and low SNR. For $N_t=32$ antennas and $0\,\si{\decibel}$ SNR, passive HPF becomes the most efficient solution as it provides enough data throughput and its PSN does not dissipate too much power due to the comparatively small number of phase-shifting elements. As the number $N_t$ of antennas increase, however, its energy efficiency quickly drops. These results also suggest that  asymptotically expanding the transmit array size does not improve energy efficiency, as static power consumption becomes asymptotically large with $N_t$ for fixed SNR, while spectral efficiency does not grow as fast.
		
	~
	~
	~
	
	\section{Conclusion} \label{sec:conc}
	
	MmWave Massive MIMO systems using hybrid precoding with fully-connected PSN exhibit important power losses mostly due to the large number of phase-shifters, power dividers and combiners. Such losses are not easily compensated for with present mmWave PA technology, causing severe spectral and energy efficiency degradation. As an alternative, hybrid precoders equipped with partially-connected PSN are shown to be energy-efficient, as they consume less power and exhibit limited insertion loss. Our results revealed that phase-shifting topology offers an energy-spectral efficiency trade-off. Active phase-shifters favor higher data throughput, while passive elements aim at energy efficiency. Moreover, we observed that DACs with more than $3$ bits of resolution do not significantly improve spectral efficiency, and may lead to energy efficiency degradation. Finally, the proposed power consumption model suggests that computational power spending should be considered in the power budget of very large array systems. The present work considered a point-to-point MIMO system for simplicity reasons. Introducing interfering users, performing transceiver optimization, and considering imperfect channel state information are envisioned as future work. 

	\appendix
	
	We demonstrate that the power of the lossless transmitted signal with DAC approximation \eqref{eq:precod_lowres} follows the average power constraint for hybrid precoding. This power is defined as $P_{\tilde{x}} = \esp{\| \tilde{\mb{x}} \|_2^2 } =  \tr(\Frf \Ups \Fbb \Fbb^\hermit \Ups \Frf^\hermit) + \tr(\Frf \mb{R}_{ee}\Frf^\hermit)$. From Equation \eqref{eq:cov_quant_error_lowres}, it follows that $P_{\tilde{x}} $ can be rewritten as 
	\begin{equation}
		P_{\tilde{x}} \approx (1-\rho_b) \frob{\Frf \Fbb}^2 + \rho_b \tr(\Frf \diag(\Fbb \Fbb^\hermit)\Frf^\hermit).
	\end{equation}
	We need to calculate the diagonal matrix $\diag(\Fbb \Fbb^\hermit)$. To do so, consider $\mb{A} = \theta \mb{I}_M$. One can readily see that $\diag(\mb{A}) = \Diag([\tr(\mb{A})/M, \ldots, \tr(\mb{A})/M])$. Although the diagonal elements of $\Fbb\Fbb^\hermit$ are not necessarily equal, they should be similar due to power allocation, yielding the following approximation $\diag(\Fbb \Fbb^\hermit) \approx \Diag(\tr(\Fbb \Fbb^\hermit)/L_t, \ldots, \tr(\Fbb \Fbb^\hermit)/L_t) \approx \frob{\Fbb}^2/L_t$. This estimate gives 
	\begin{equation} 
	P_{\tilde{x}}\approx (1-\rho_b)\frob{\Frf \Fbb}^2 + \rho_b \frac{\frob{\Fbb}^2 \frob{\Frf}^2}{L_t}.
	\end{equation} 
	The baseband precoder normalization \eqref{eq:baseband} leads to $\frob{\Fbb}^2 = \frac{\Pmax}{\frob{\Frf \mb{Q}}^2}\tr(\mb{Q} \mb{Q}^\hermit)$. From water-filling power allocation, $\tr(\mb{Q} \mb{Q}^\hermit) = \Pmax$, and the denominator can be written as $\frob{\Frf \mb{Q}}^2 = \tr(\Frf \mb{V} \wf \mb{V}^\hermit \Frf^\hermit)$. Notice that we can approximate $\mb{Q}\mb{Q}^\hermit \approx \frac{\Pmax}{L_t}\mb{I}_{L_t}$, therefore 
	\begin{equation}
	\frob{\Frf \mb{Q}}^2 \approx \frac{\Pmax}{L_t} \tr(\Frf \Frf^\hermit) = N_t \Pmax.
	\end{equation} 
	We finally get $\frob{\Fbb}^2 = \Pmax/N_t$ and $P_{\tilde{x}} \approx [(1-\rho_b) + \rho_b] \Pmax = \Pmax$. This result can be easily extended to digital precoding by making $\Fbb \in \mbb{C}^{N_t \times N_s}$ and $\Frf = \mb{I}_{N_t}$. 

	\section*{Acknowledgments}
	The authors would like to thank Prof. Josef A. Nossek and the referees for the helpful comments that contributed to the improvement of this work.

	\bibliographystyle{IEEEtran}
	\bibliography{hb}

\begin{thebibliography}{10}
\providecommand{\url}[1]{#1}
\csname url@samestyle\endcsname
\providecommand{\newblock}{\relax}
\providecommand{\bibinfo}[2]{#2}
\providecommand{\BIBentrySTDinterwordspacing}{\spaceskip=0pt\relax}
\providecommand{\BIBentryALTinterwordstretchfactor}{4}
\providecommand{\BIBentryALTinterwordspacing}{\spaceskip=\fontdimen2\font plus
\BIBentryALTinterwordstretchfactor\fontdimen3\font minus
  \fontdimen4\font\relax}
\providecommand{\BIBforeignlanguage}[2]{{%
\expandafter\ifx\csname l@#1\endcsname\relax
\typeout{** WARNING: IEEEtran.bst: No hyphenation pattern has been}%
\typeout{** loaded for the language `#1'. Using the pattern for}%
\typeout{** the default language instead.}%
\else
\language=\csname l@#1\endcsname
\fi
#2}}
\providecommand{\BIBdecl}{\relax}
\BIBdecl

\bibitem{wang_cellular_2014}
C.-X. Wang, F.~Haider, X.~Gao, X.-H. You, Y.~Yang, D.~Yuan, H.~Aggoune,
  H.~Haas, S.~Fletcher, and E.~Hepsaydir, ``Cellular architecture and key
  technologies for {5G} wireless communication networks,'' \emph{IEEE
  Communications Magazine}, vol.~52, no.~2, pp. 122--130, Feb. 2014.

\bibitem{andrews_what_2014}
J.~G. Andrews, S.~Buzzi, W.~Choi, S.~V. Hanly, A.~Lozano, A.~C.~K. Soong, and
  J.~C. Zhang, ``What will {5G} be?'' \emph{IEEE Journal on Selected Areas in
  Communications}, vol.~32, no.~6, pp. 1065--1082, Jun. 2014.

\bibitem{constantine2005antenna}
A.~B. Constantine, \emph{Antenna {T}heory: {A}nalysis and {D}esign},
  3rd~ed.\hskip 1em plus 0.5em minus 0.4em\relax Wiley, 2005.

\bibitem{heath_overview_2016}
R.~W. Heath, N.~Gonzalez-Prelcic, S.~Rangan, W.~Roh, and A.~M. Sayeed, ``An
  overview of signal processing techniques for millimeter wave {MIMO}
  systems,'' \emph{IEEE Journal of Selected Topics in Signal Processing},
  vol.~10, no.~3, pp. 436--453, Apr. 2016.

\bibitem{larsson_massive_2014}
E.~G. Larsson, O.~Edfors, F.~Tufvesson, and T.~L. Marzetta, ``Massive {MIMO}
  for next generation wireless systems,'' \emph{IEEE Communications Magazine},
  vol.~52, no.~2, pp. 186--195, Feb. 2014.

\bibitem{schwarz_society_2016}
S.~Schwarz and M.~Rupp, ``Society in motion: challenges for {LTE} and beyond
  mobile communications,'' \emph{IEEE Communications Magazine}, vol.~54, no.~5,
  pp. 76--83, May 2016.

\bibitem{heath_millimeter_2016}
R.~W. Heath, ``Millimeter wave: the future of commercial wireless systems,'' in
  \emph{Proc. 2016 {IEEE} {Compound} {Semiconductor} {Integrated} {Circuit}
  {Symposium} ({CSICS})}, Austin, TX, USA, Oct. 2016, pp. 1--4.

\bibitem{el_ayach_spatially_2014}
O.~El~Ayach, S.~Rajagopal, S.~Abu-Surra, Z.~Pi, and R.~W. Heath, ``Spatially
  sparse precoding in millimeter wave {MIMO} systems,'' \emph{IEEE Transactions
  on Wireless Communications}, vol.~13, no.~3, pp. 1499--1513, Mar. 2014.

\bibitem{mendez-rial_hybrid_2016}
R.~M\'endez-Rial, C.~Rusu, N.~González-Prelcic, A.~Alkhateeb, and R.~W. Heath,
  ``Hybrid {MIMO} architectures for millimeter wave communications: {Phase}
  shifters or switches?'' \emph{IEEE Access}, vol.~4, pp. 247--267, 2016.

\bibitem{tsinos_energy-efficiency_2017}
C.~Tsinos, S.~Maleki, S.~Chatzinotas, and B.~Ottersten, ``On the
  energy-efficiency of hybrid analog-digital transceivers for single- and
  multi-carrier large antenna array systems,'' \emph{IEEE Journal on Selected
  Areas in Communications}, vol.~35, no.~9, pp. 1980--1995, Sep. 2017.

\bibitem{garcia-rodriguez_hybrid_2016}
A.~Garcia-Rodriguez, V.~Venkateswaran, P.~Rulikowski, and C.~Masouros, ``Hybrid
  analog-digital precoding revisited under realistic {RF} modeling,''
  \emph{IEEE Wireless Communications Letters}, vol.~5, no.~5, pp. 528--531,
  Oct. 2016.

\bibitem{orhan_low_2015}
O.~Orhan, E.~Erkip, and S.~Rangan, ``Low power analog-to-digital conversion in
  millimeter wave systems: {Impact} of resolution and bandwidth on
  performance,'' in \emph{Proc. 2015 Information {Theory} and {Applications}
  {Workshop} ({ITA})}, San Diego, CA, USA, Feb. 2015, pp. 191--198.

\bibitem{abbas_millimeter_2016}
W.~B. Abbas, F.~Gomez-Cuba, and M.~Zorzi, ``Millimeter wave receiver
  efficiency: a comprehensive comparison of beamforming schemes with low
  resolution {ADCs},'' \emph{IEEE Transactions on Wireless Communications},
  vol.~16, no.~12, pp. 8131--8146, Dec. 2017.

\bibitem{tropp_designing_2005}
J.~A. Tropp, I.~S. Dhillon, R.~W. Heath, and T.~Strohmer, ``Designing
  structured tight frames via an alternating projection method,'' \emph{IEEE
  Transactions on Information Theory}, vol.~51, no.~1, pp. 188--209, Jan. 2005.

\bibitem{roth_achievable_2016}
K.~Roth and J.~A. Nossek, ``Achievable rate and energy efficiency of hybrid and
  digital beamforming receivers with low resolution {ADC},'' \emph{IEEE Journal
  on Selected Areas in Communications}, vol.~35, no.~9, pp. 2056--2068, Sep.
  2017.

\bibitem{marzetta_fundamentals_2016}
T.~L. Marzetta, E.~G. Larsson, H.~Yang, and H.~Q. Ngo, \emph{Fundamentals of
  {Massive} {MIMO}}.\hskip 1em plus 0.5em minus 0.4em\relax Cambridge
  University Press, 2016.

\bibitem{ribeiro_low-complexity_2017}
L.~N. Ribeiro, S.~Schwarz, M.~Rupp, A.~L.~F. de~Almeida, and J.~C.~M. Mota, ``A
  {Low}-{Complexity} {Equalizer} for {Massive} {MIMO} {Systems} {Based} on
  {Array} {Separability},'' in \emph{Proc. 25th {European} {Signal}
  {Processing} {Conference} (EUSIPCO)}, Kos, Greece, Aug. 2017, pp. 2522--2526.

\bibitem{mezghani_transmit_2009}
A.~Mezghani, R.~Ghiat, and J.~A. Nossek, ``Transmit processing with low
  resolution {D}/{A}-converters,'' in \emph{Proc. 16th {IEEE} {International}
  {Conference} on {Electronics}, {Circuits}, and {Systems} ({ICECS})}, Yasmine
  Hammamet, Tunisia, Dec. 2009, pp. 683--686.

\bibitem{bussgang_crosscorrelation_1952}
\BIBentryALTinterwordspacing
J.~J. Bussgang, ``Crosscorrelation functions of amplitude-distorted {Gaussian}
  signals,'' Research Laboratory of Electronics, Massachusetts Institute of
  Technology, Tech. Rep. 216, Mar. 1952. [Online]. Available:
  \url{https://dspace.mit.edu/handle/1721.1/4847}
\BIBentrySTDinterwordspacing

\bibitem{jacobsson_quantized_2016}
S.~Jacobsson, G.~Durisi, M.~Coldrey, T.~Goldstein, and C.~Studer, ``Quantized
  precoding for massive {MU}-{MIMO},'' \emph{IEEE Transactions on
  Communications}, vol.~65, no.~11, pp. 4670--4684, Nov. 2017.

\bibitem{bjornson_optimal_2015}
E.~Bj\"ornson, L.~Sanguinetti, J.~Hoydis, and M.~Debbah, ``Optimal design of
  energy-efficient multi-user {MIMO} systems: {Is} massive {MIMO} the answer?''
  \emph{IEEE Transactions on Wireless Communications}, vol.~14, no.~6, pp.
  3059--3075, Jun. 2015.

\bibitem{golub_matrix_2012}
G.~H. Golub and C.~F. Van~Loan, \emph{Matrix {C}omputations}, 3rd~ed.\hskip 1em
  plus 0.5em minus 0.4em\relax JHU Press, 2012.

\bibitem{sohrabi_hybrid_2015}
F.~Sohrabi and W.~Yu, ``Hybrid beamforming with finite-resolution phase
  shifters for large-scale {MIMO} systems,'' in \emph{Proc. {IEEE} 16th
  {International} {Workshop} on {Signal} {Processing} {Advances} in {Wireless}
  {Communications} ({SPAWC})}, Stockholm, Sweden, Jul. 2015, pp. 136--140.

\bibitem{max_quantizing_1960}
J.~Max, ``Quantizing for minimum distortion,'' \emph{IRE Transactions on
  Information Theory}, vol.~6, no.~1, pp. 7--12, Mar. 1960.

\bibitem{makhoul_vector_1985}
J.~Makhoul, S.~Roucos, and H.~Gish, ``Vector quantization in speech coding,''
  \emph{Proceedings of the IEEE}, vol.~73, no.~11, pp. 1551--1588, Nov. 1985.

\bibitem{mezghani_capacity_2012}
A.~Mezghani and J.~A. Nossek, ``Capacity lower bound of {MIMO} channels with
  output quantization and correlated noise,'' in \emph{Proc. {IEEE}
  {International} {Symposium} on {Information} {Theory} {Proceedings}
  ({ISIT})}, Cambridge, MA, USA, Jul. 2012.

\bibitem{diggavi_worst_2001}
S.~N. Diggavi and T.~M. Cover, ``The worst additive noise under a covariance
  constraint,'' \emph{IEEE Transactions on Information Theory}, vol.~47, no.~7,
  pp. 3072--3081, Nov. 2001.

\bibitem{yu_iterative_2004}
W.~Yu, W.~Rhee, S.~Boyd, and J.~M. Cioffi, ``Iterative water-filling for
  {Gaussian} vector multiple-access channels,'' \emph{IEEE Transactions on
  Information Theory}, vol.~50, no.~1, pp. 145--152, Jan. 2004.

\bibitem{gao_mmwave_2015}
Z.~Gao, L.~Dai, D.~Mi, Z.~Wang, M.~A. Imran, and M.~Z. Shakir, ``{mmWave}
  massive-{MIMO}-based wireless backhaul for the {5G} ultra-dense network,''
  \emph{IEEE Wireless Communications}, vol.~22, no.~5, pp. 13--21, Oct. 2015.

\bibitem{zochmann_directional_2016}
E.~Z\"ochmann, M.~Lerch, S.~Caban, C.~F. Mecklenbr\"auker, and M.~Rupp,
  ``Directional evaluation of receive power, {Rician} {K}-factor and {RMS}
  delay spread obtained from power measurements of 60 {GHz} indoor channels,''
  in \emph{Proc. {IEEE} {APS} {Topical} {Conference} on {Antennas} and
  {Propagation} in {Wireless} {Communications} ({APWC})}, Cairns, Australia,
  2016, pp. 246--249.

\bibitem{huang_60_2017}
D.~Huang, L.~Zhang, L.~Zhang, and Y.~Wang, ``A 60 {GHz} {$360^\circ$} 5-bit
  digitally controlled high-pass/direct-pass {T}-type phase shifter using
  {nMOS} body-floating switch,'' \emph{Analog Integrated Circuits and Signal
  Processing}, vol.~90, no.~1, pp. 93--100, Jan. 2017.

\bibitem{prasad_energy_2017}
K.~S.~V. Prasad, E.~Hossain, and V.~K. Bhargava, ``Energy {efficiency} in
  {massive} {MIMO}-{based} {5G} {networks}: {Opportunities} and {challenges},''
  \emph{IEEE Wireless Communications}, vol.~24, no.~3, pp. 86--94, Jun. 2017.

\bibitem{marcu_lo_2011}
\BIBentryALTinterwordspacing
C.~Marcu, ``{LO} {Generation} and {Distribution} for 60 {GHz} {Phased} {Array}
  {Transceivers},'' Ph.D. dissertation, EECS Department, University of
  California, Berkeley, 2011. [Online]. Available:
  \url{http://www2.eecs.berkeley.edu/Pubs/TechRpts/2011/EECS-2011-132.html}
\BIBentrySTDinterwordspacing

\bibitem{jin_7_2011}
Y.~Jin, J.~R. Long, and M.~Spirito, ``A 7 {dB} {NF} 60 {GHz}-band
  millimeter-wave transconductance mixer,'' in \emph{Proc. 2011 {IEEE} {Radio}
  {Frequency} {Integrated} {Circuits} {Symposium} ({RFIC})}, Baltimore, MD,
  USA, Jun. 2011, pp. 1--4.

\bibitem{rangan_energy_2013}
S.~Rangan, T.~Rappaport, E.~Erkip, Z.~Latinovic, M.~R. Akdeniz, and Y.~Liu,
  ``Energy efficient methods for millimeter wave picocellular systems,'' in
  \emph{Proc. {IEEE} {Communications} {Theory} {Workshop}}, Phuket, Thailand,
  Jun. 2013, pp. 1--25.

\bibitem{scheir_52_2008}
K.~Scheir, S.~Bronckers, J.~Borremans, P.~Wambacq, and Y.~Rolain, ``A 52 {GHz}
  phased-array receiver front-end in 90 nm digital {CMOS},'' \emph{IEEE Journal
  of Solid-State Circuits}, vol.~43, no.~12, pp. 2651--2659, Dec. 2008.

\bibitem{cui_energy-constrained_2005}
S.~Cui, A.~J. Goldsmith, and A.~Bahai, ``Energy-constrained modulation
  optimization,'' \emph{IEEE Transactions on Wireless Communications}, vol.~4,
  no.~5, pp. 2349--2360, Sep. 2005.

\bibitem{olieman_time-interleaved_2016}
\BIBentryALTinterwordspacing
E.~Olieman, ``Time-interleaved high-speed {D}/{A} converters,'' Ph.D.
  dissertation, University of Twente, 2016. [Online]. Available:
  \url{http://doc.utwente.nl/99674/}
\BIBentrySTDinterwordspacing

\bibitem{gao_power-performance_2017}
K.~Gao, N.~J. Estes, B.~Hochwald, J.~Chisum, and J.~N. Laneman,
  ``Power-performance analysis of a simple one-bit transceiver,'' in
  \emph{Proc. 2017 {Information} {Theory} and {Applications} {Workshop}
  ({ITA})}, San Diego, CA, USA, Feb. 2017, pp. 1--10.

\bibitem{greene_60-ghz_2017}
K.~Greene, A.~Sarkar, and B.~Floyd, ``A 60-{GHz} {Dual}-{Vector} {Doherty}
  {Beamformer},'' \emph{IEEE Journal of Solid-State Circuits}, vol.~52, no.~5,
  pp. 1373--1387, May 2017.

\bibitem{pepe_78.8-92.8_2015}
D.~Pepe and D.~Zito, ``A {$78.8$}-{$92.8$} {GHz} {$4$}-bit {$0$}-{$360^\circ$}
  active phase shifter in 28nm {FDSOI} {CMOS} with 2.3 {dB} average peak
  gain,'' in \emph{Proc. 41st {European} {Solid}-{State} {Circuits}
  {Conference} ({ESSCIRC})}, Graz, Austria, Sep. 2015, pp. 64--67.

\bibitem{biglieri_mimo_2007}
E.~Biglieri, R.~Calderbank, A.~Constantinides, A.~Goldsmith, A.~Paulraj, and
  H.~V. Poor, \emph{{MIMO} {Wireless} {Communications}}.\hskip 1em plus 0.5em
  minus 0.4em\relax Cambridge University press, 2007.

\bibitem{zochmann_comparing_2016}
E.~Z\"ochmann, S.~Schwarz, and M.~Rupp, ``Comparing antenna selection and
  hybrid precoding for millimeter wave wireless communications,'' in
  \emph{Proc. 2016 {IEEE} {Sensor} {Array} and {Multichannel} {Signal}
  {Processing} {Workshop} ({SAM})}, Rio de Janeiro, Brazil, Jul. 2016, pp.
  1--5.

\end{thebibliography}
	
\begin{IEEEbiography}[{\includegraphics[width=1in,height=1.25in,clip,keepaspectratio]{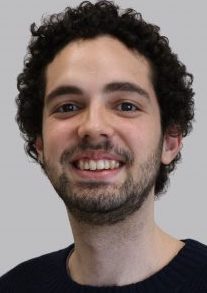}}]{Lucas N. Ribeiro}
	received his Bachelor degree in Teleinformatics Engineering from the Federal University of Cear\'a, Brazil, in 2014, and his Master degree in Informatics from the University of Nice Sophia Antipolis, France, in 2015. He is currently pursuing the Ph.D. degree in Teleinformatics Engineering in the Wireless Telecommunications Research Group, Federal University of Cear\'a. In 2017, he was a visiting researcher at the Institute of Telecommunications of Technische Universit\"at Wien.
\end{IEEEbiography}
~
\begin{IEEEbiography}[{\includegraphics[width=1in,height=1.25in,clip,keepaspectratio]{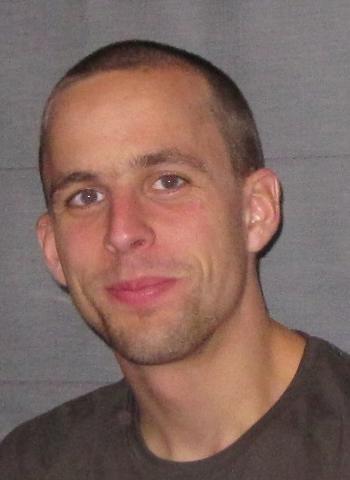}}]{Stefan Schwarz}
received his Dr. techn. degree in telecommunications engineering in 2013 at Technische Universit\"at (TU) Wien. In 2010 he received the honorary price of the Austrian Minister of Science and Research and in 2014 he received the INiTS ICT award. From 2008 to 2014 he was working at the Institute of Telecommunications (ITC) of TU Wien. Since 2015 he is heading the Christian Doppler Laboratory for Dependable Wireless Connectivity for the Society in Motion.
\end{IEEEbiography}
~
\begin{IEEEbiography}[{\includegraphics[width=1in,height=1.25in,clip,keepaspectratio]{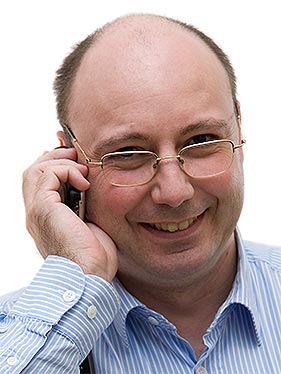}}]{Markus Rupp}
received his Dr.-Ing. degree in 1993 at Technische Universit\"at Darmstadt. From 1993 until 1995, he had a postdoctoral position at the University of Santa Barbara, California. From October 1995 until August 2001 he was a member of Technical Staff in the Wireless Technology Research Department of Bell-Labs at Crawford Hill, NJ. Since October 2001 he is a full professor for Digital Signal Processing in Mobile Communications at Technische Universit\"at Wien.
\end{IEEEbiography}
~
\begin{IEEEbiography}[{\includegraphics[width=1in,height=1.25in,clip,keepaspectratio]{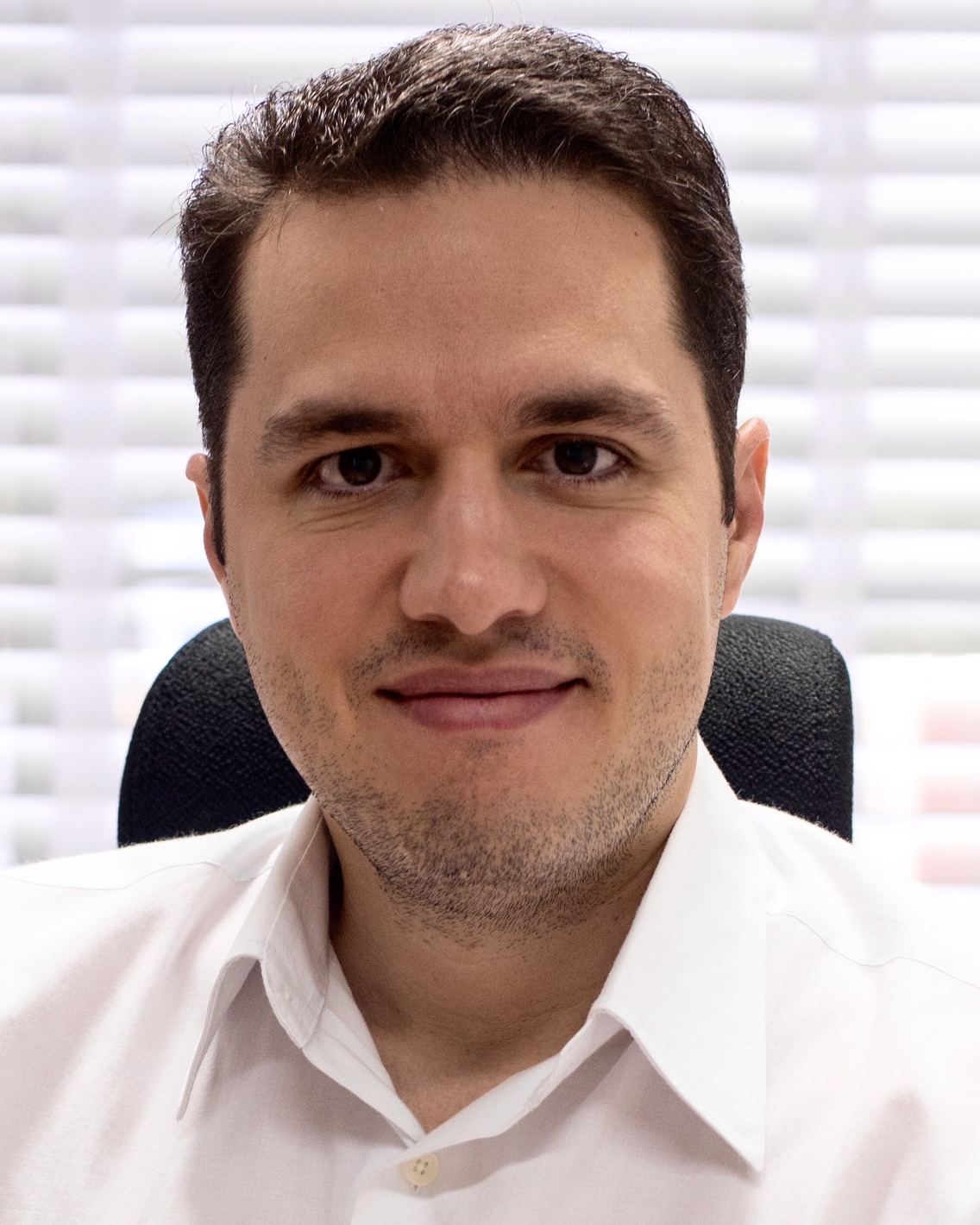}}]{Andr\'e L. F. de Almeida} received the B.Sc. and M.Sc. degrees in Electrical Engineering from the Federal University of Cear\'a, Brazil, in 2001 and 2003, respectively, and the double Ph.D. degree in Sciences and Teleinformatics Engineering from the University of Nice Sophia Antipolis, France, and the Federal University of Cear\'a, Brazil, in 2007. He is currently an Associate Professor with the Department of Teleinformatics Engineering of the Federal University of Cear\'a. 
\end{IEEEbiography}
	
\end{document}